# Terahertz photocurrent probe of quantum geometry and interactions in magic-angle twisted bilayer graphene

+R. Krishna Kumar[1], *G. Li[1], *R. Bertini[1], *S. Chaudhary[2,3], K. Nowakowski[1], J. M. Park[3], S. Castilla[1], Z. Zhan[4], P. A. Pantaleón[4], H. Agarwal[1], S. Battle-Porro[1], E. Icking[5,6], M. Ceccanti[1], Antoine Reserbat-Plantey[1,7], G. Piccinini[1], J. Barrier[1], E. Khestanova[1], T. Taniguchi[8], K. Watanabe[9], C. Stampfer[5,6], G. Refael[10], F. Guinea[4,11], P. Jarillo-Herrero[3], J. C. W Song[12], P. Stepanov[1,13,14], C. Lewandowski[10,15,16], +F. H. L. Koppens[1,17]

1. ICFO - Institut de Ciencies Fotoniques, The Barcelona Institute of Science and Technology, 08860 Castelldefels (Barcelona), Spain.
2. Department of Physics, The University of Texas, Austin, TX 78712, USA.
3. Department of Physics, Massachusetts Institute of Technology, Cambridge, MA 02139, USA.
4. Imdea Nanoscience, Faraday 9, 28049, Madrid, Spain
5. JARA-FIT and 2nd Institute of Physics, RWTH Aachen University, 52074 Aachen, Germany, EU
6. Peter Grünberg Institute (PGI-9), Forschungszentrum Jülich, 52425 Jülich, Germany, EU
7. Université Côte d'Azur, CNRS, CRHEA. Valbonne, Sophia-Antipolis, France
8. Research Center for Materials Nanoarchitectonics, National Institute for Materials Science, 1-1 Namiki, Tsukuba 305-0044, Japan
9. Research Center for Electronic and Optical Materials, National Institute for Materials Science, 1-1 Namiki, Tsukuba 305-0044, Japan
10. Department of Physics and Institute for Quantum Information and Matter, California Institute of Technology, Pasadena, California 91125, USA
11. Donostia International Physics Center, Paseo Manuel de Lardizabal 4, 20018 San Sebastian, Spain
12. Division of Physics and Applied Physics, School of Physical and Mathematical Sciences, Nanyang Technological University, Singapore, 637371, Singapore
13. Department of Physics and Astronomy, University of Notre Dame, Notre Dame, Indiana 46556, USA
14. Stavropoulos Center for Complex Quantum Matter, University of Notre Dame, Notre Dame, IN 46556, USA
15. National High Magnetic Field Laboratory, Tallahassee, Florida, 32310, USA
16. Department of Physics, Florida State University, Tallahassee, Florida 32306, USA
17. ICREA-Institució Catalana de Recerca i Estudis Avançats, 08010 Barcelona, Spain

*These authors contributed equally
+ corresponding authors: roshan.krishnakumar@icfo.eu, frank.koppens@icfo.eu

**Moiré materials represent strongly interacting electron systems bridging topological and correlated physics. Despite significant advances, decoding wavefunction properties underlying the quantum geometry remains challenging. Here, we utilize polarization-resolved photocurrent measurements to probe magic-angle twisted bilayer graphene, leveraging its sensitivity to the Berry connection that encompasses quantum "textures" of electron wavefunctions. Using terahertz light resonant with optical transitions of its flat bands, we observe bulk photocurrents driven by broken symmetries and reveal the interplay between electron interactions and quantum geometry. We observe inversion-breaking gapped states undetectable through quantum transport, sharp changes in the polarization axes caused by interaction-induced band renormalization, and recurring photocurrent patterns at integer fillings of the moiré unit cell that track the evolution of quantum geometry through the cascade of phase transitions. The large and tunable terahertz response intrinsic to flat-band systems offers direct insights into the quantum geometry of interacting electrons and paves the way for innovative terahertz quantum technologies.**

Twisted bilayer graphene (TBG) represents a new paradigm for exploring the interplay of strongly interacting phenomena[1,2,3] with non-trivial topology[4,5,6]. While previous experiments have revealed several exotic correlated and topological phenomena[7], measurements of quantum geometry remain challenging, relying on the observation of broken-symmetry phases driven by electron-electron interactions[4,5,6] that occur only at specific band fillings. Furthermore, the underlying symmetries driving such phenomena are often local[8,9] because of their sensitivity to substrates[10], strain[11], and reconstruction[12] making them easily smeared in global measurements. Photocurrent measurements provide a powerful complementary probe to studying quantum materials[13]. The bulk photovoltaic effect (BPE)[13,14,15] represents an intriguing class of photocurrent phenomena where the current is generated by broken symmetries intrinsic to the crystal. Aside from its promise for photovoltaics[16], it has gained significant attention for studying quantum materials because of its ties to the second-order conductivity, whose magnitude is controlled by geometric properties of electron wavefunctions and the projection of electric field ($E$) relative to the crystal lattice $J = \sigma^c_{ab}E_aE_b$. Moreover, even if photoexcitation is global, local photoactive regions of interest may still be discerned, making the bulk photovoltaic effect a potentially more sensitive probe of electronic orders.

Polarization-dependent measurements allow to rotate $E$ arbitrarily (Fig. 1a) which serves as a powerful tool for probing quantum geometry and underlying symmetries of the system. The polarization phase ($\alpha$), the angle of polarization ($\theta_P$) relative to contacts where the current is maximum (Fig. 1b), describes an intrinsic quantum direction manifested by geometric phases accumulated when electrons transition within (intraband) or between (interband) manifolds of quantum states (Fig. 1c). While distinct microscopic mechanisms exist coupled to different physical quantities[13], e.g. Berry curvature or shift-vector, they are all built from various gauge-invariant combinations of Berry connection, which captures changes in the internal structure of electron wavefunctions between two quantum mechanical states coupled by optical fields. Therefore, in moiré materials second-order conductivities are anticipated to be extraordinarily large[17,18,19,20,21] because their giant superlattice unit cell ($\sim$10 nm) sets the scale of the Berry connection[22]. Moreover, they can be particularly strong at terahertz (THz) frequencies because of the large joint density of states resonant at millielectronvolt (meV) energy scales inherent to flat-bands[20,19] (Inset, Fig. 1e). Here, we report a large and polarization-sensitive THz photoresponse in magic-angle TBG and use it to study the interplay between electron interactions and quantum geometry at the energy scale of its flat-bands. Specifically, we probe the quantum geometric textures of the gapped state at charge neutrality[23], electron-electron interaction-induced band renormalizations[24], and cascade phenomena[25,26].

**Results**

We studied four devices close to the magic-angle with twist angles of 0.94°, 1.02°, 1.03° and 1.12°, denoted as D0.94, D1.02, D1.03, D1.12, respectively (see methods for fabrication). All devices showed oscillating features in the resistivity around integer fillings at 10 K, which we attribute to the parent states of correlated phenomena (Fig. 1d/Fig. S8a). However, their behaviour differs at the charge neutrality point (CNP). D0.94 and D1.12 showed a monotonic increase in resistivity with temperature, indicating metallic behaviour, while D1.03 exhibits a marked insulating peak showing activation behaviour upon increasing temperature (Fig. 1d inset). This behaviour indicates gapping of the flat bands, as expected for TBG aligned to hexagonal boron nitride (hBN)[10]. The gap serves as a signature of broken $C_{2z}T$ symmetry, where $C_{2z}$ denotes a two-fold rotation about the z-axis and T time-reversal, which is a prerequisite for driving in-plane bulk photocurrents[14]. Nonetheless, in all our devices we observed a strong bulk THz photoresponse that is polarization-dependent, with similar characteristics as discussed below.

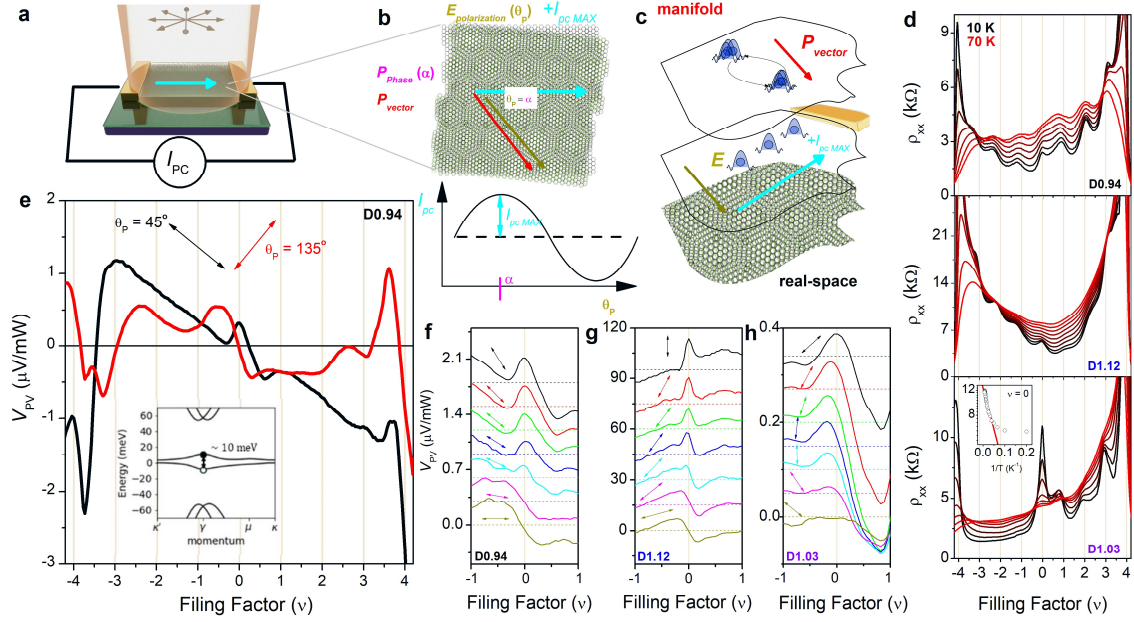

**Figure. 1 Polarization-dependent terahertz photocurrents in twisted bilayer graphene. a,** Schematic of terahertz photocurrent experiments. Yellow arrows indicate the polarization of light. Solid cyan arrows indicate the projection of the generated photocurrent along the measurement contacts ($I_{PC}$). **b,** Top Panel: Schematic of the underlying TBG (grey) on hBN (green). Cyan, red and dark yellow arrows depict $I_{PC}$, the polarization vector intrinsic to quantum geometry, and the polarization direction ($\theta_P$) of the incident electric field respectively. When the polarization of light is aligned with the polarization vector, the photocurrent is maximum. The polarization vector forms an angle with the current projection called the polarization phase ($\alpha$). Bottom panel: $I_{PC}(\theta_P)$ is sinusoidal with a maximum for $\theta_P = \alpha$. **c,** schematic of quantum geometric response. The top slice depicts electrons transitioning between quantum states driven by the optical field. Electric field alignment with the polarization vector drives an anomalous current response (middle slice). **d,** Resistivity ($\rho_{xx}$) as a function of filling factor ($\nu$) plotted for different temperatures in D0.94, D1.12 and D1.03. Inset bottom panel: $\rho_{xx}$ (T) for $\nu = 0$ (open circles). Red dashed line plots an exponential fit from which gaps of 3.6 meV are extracted. **e,** Photovoltage ($V_{PV}$) as a function $\nu$, plotted for two orthogonal polarizations of terahertz light (black and red curves). Inset: band structure of magic-angle TBG with terahertz transitions. **f,** zoomed data of $V_{PV}$ close to the CNP for different polarization directions. Solid lines are experimental data offset for clarity and the dashed lines indicate the $V_{PV} = 0$ line. **g,h** same as **f** for D1.12 and D1.03 respectively. Data from D1.12 was performed at 0.7 THz, (2.7 meV excitation) whilst all other data sets were performed at 2.5 THz (10 meV).

**Polarization-resolved terahertz photocurrent measurements**

Figure 1e displays the photovoltage ($V_{PV}$) as a function of filling factor ($\nu$) measured in D0.94 for two orthogonal polarizations normalized by the incident power. We find the photoresponse for the two directions is strikingly different. The response differs not only in magnitude, the $\nu$ dependence also changes dramatically even changing sign. The functional form also contrasts greatly with the photothermoelectric effect[27] (Supplementary Section 1), pointing towards a distinct origin. Of particular significance is the behaviour around the CNP, which exhibits a peak for one polarization and zero-crossing behaviour for the other. Figure 1f details this behaviour, showing the peaked response evolving smoothly from the sign-changing response with polarization. Similar behaviour was observed for D1.12 (Fig. 1g) and D1.03 (Fig. 1h): For full dependence, see Supplementary Section 2. D1.12 required smaller excitation energies to reveal such behaviour (Supplementary Section 3), we attribute to the excitation energy approaching the inter-flat band resonances of magic-angle TBG (Fig. 1e inset). Figure 1 shows that the photoresponse around the CNP is very sensitive to polarization.

Figure 2a plots the full polarization-dependent response $V_{PV}(\theta_p)$ for different $\nu$ close to CNP. For all $\nu$, we observe an oscillating photoresponse indicating the effect originates from light polarization (See Supplementary Section 4 for additional checks). More notably, we find that the polarization phase ($\alpha$), is highly sensitive to $\nu$. This behaviour contrasts greatly with usual junction phenomena[27], device geometry[28] and antenna[29,30] effects, where $\alpha$ would be $\nu$ independent (see Supplementary Section 5). Hence, the $\nu$ dependence underscores that the photocurrents and their polarization dependence are primarily governed by the electronic system.

To further analyse changes in $\alpha$, we isolate the polarization-dependent component of the photocurrent ($V_{Lin}$) by fitting a sinusoidal function (Fig. 2a) and subtracting the polarization-independent background. Figure 2b plots $V_{Lin}(\theta_p,\nu)$. The white lines mark sign changes in the photoresponse, thereby tracking the changes in $\alpha$. Strikingly, we find a strong change with $\nu$. A drift in $\alpha$ is observed at the CNP (raw data in Fig. 2a) and additional sharp drifts around $\nu$ = -1 (Fig. 2c). Figure 2d plots $\alpha(\nu)$ highlighting the winding of $\alpha$ at the CNP and integer $\nu$. $\alpha$ stabilizes for intermediate $\nu$ and then drifts again approaching $\nu = \pm 4$. Significant changes were also observed in D1.03 (Fig. 2e) and D1.12 (Fig. 2f) (see Supplementary Section 2), although the specifics vary across different devices. This variation reflects subtle differences in electronic structure and competing contributions to the measured photoresponse. Key features, particularly sharp drifts in $\alpha$ close to the CNP and $\nu = \pm$ 4, were identified in all the studied devices. Moreover, features near integer fillings appear as peaks/dips in $\alpha$ (indicated by red arrows) or sharp drifts in $\alpha$ (blue arrows) resembling the behaviour at the CNP.

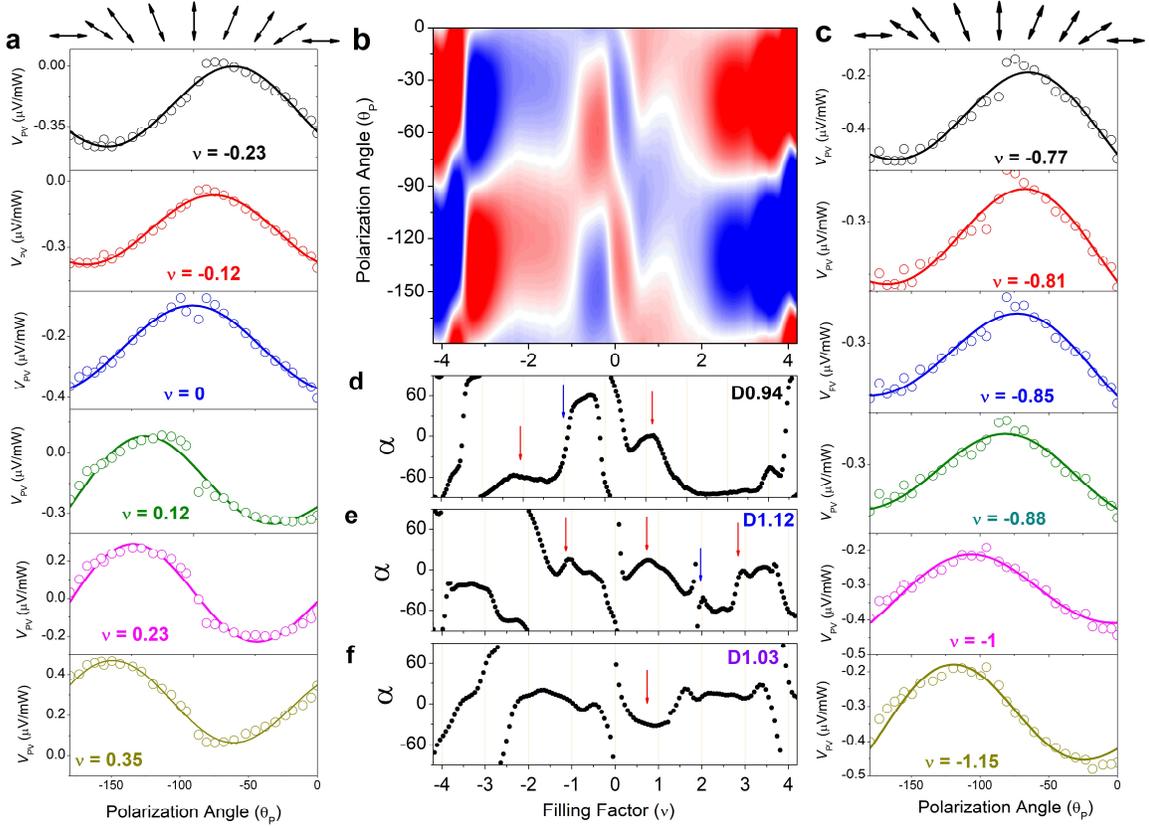

**Figure. 2. Filling Factor dependence of the polarization phase in twisted bilayer graphene. a,** Photovoltage ($V_{PV}$) measured as a function of polarization direction ($\theta_P$) for different fillings (top to bottom) in D 0.94. Open circles are experimental data and solid lines plot fits to equation (1). **b,** Polarization dependent component of the photoresponse ($V_{Lin}$) as a function of ν and $\theta_P$. Colour scale: blue: red, -0.5:0.5 μV/mW. **c,** same as **a**, for filing close to ν = -1. **d,** polarization phase (α) as a function of filling factor (ν) for D 0.94 extracted from **b**. **e,f,** same as **d** for D1.03 and D1.12. The red and blue arrows in **d,e** and **f** highlight integer-filling features.

### Hartree-renormalized shift current response

The changes in α with ν provide strong evidence for its connection to the bulk photovoltaic effect[14,17]. In this case, the photoresponse is governed by second-order conductivities, $J_c = \sigma_{ab}^c E_a E_b$, $\{a,b,c\} = \{x,y\}$, where $E_x, E_y$ describe the components of the electric field projected onto the principal vectors of the 2D crystal lattice. We proceed to understand possible contributions and compare them with those extracted from our experiment by decomposing α into two principal vectors with amplitudes $L1$ and $L2$, according to α=tan$^{-1}$(L1/L2), which describes the photoresponse as

$$V_{PV} = L1\sin(2\theta) + L2\cos(2\theta) + D, \qquad (1)$$

where D is the polarization-independent contribution. This signal decomposition is formally identical to that in Fig. 2d-f, however, it allows a more direct comparison between conductivity tensor elements and the experimental coefficients (Supplementary Section 6). Specifically, for systems with C3 symmetry there exist two independent components, $\sigma_{xxx}/\sigma_{xxy}$, which, up to a rotation of the reference frame, correspond to L1/L2 in equation (1). The extracted coefficients are plotted in Fig. 3b for D0.94, aligning the reference frame with the direction where the CNP response is maximum ($\theta_p \sim 45°$).

Following, we must identify the microscopic mechanism of the photoresponse. We first focus on the behaviour around the CNP where all devices show photoresponses evolving from a sign-changing to a peaked response (Fig. 1f-h), alongside drifts in $\alpha$ (Fig. 2a,d-f). There are two possible contributions to second-order conductivity: intraband and interband excitations. The former, pertains to quantum nonlinear electrical responses[31,32,18]. However, measurements of the second-harmonic voltage, a typical signature of these Fermi surface effects, showed rather different behaviour distinguishing its microscopic origin (Supplementary Section 7). Crucially, intraband effects could not explain the peaked response at CNP. In the absence of a Fermi surface, this response is more characteristic of an interband transition[19,20]. As the photoresponse arises from linearly polarized light preserving time-reversal symmetry, we tentatively assign its origin to shift currents. These are second-order responses that arises from a real-space shift experienced by an electron wave function upon an interband excitation driven by linearly polarized light (Fig. 3a). These photocurrents were previously reported in moiré materials at infrared frequencies[17] and are expected to be amplified at terahertz frequencies where it is more sensitive to the flat-band physics.

To account for shift currents, we calculate the two components of the conductivities $\sigma_{xxx}$, $\sigma_{xxy}$ (Supplementary Section 6). Because our excitation energies are smaller than the flat-remote band gaps (Supplementary Section 8), we consider only transitions between flat bands (Fig. 3a). Figure 3d,e show these calculations for two scenarios: the non-interacting (Fig. 3d) and the Hartree-renormalized band structure[34] (Fig. 3e), which accounts for the charge redistribution of the TBG system[34] and is expected to be the leading interaction effect above 10 K. Comparing calculations (Fig. 3d,e left) with measured *L1, L2* (Fig. 3c, left), we find our photoresponse is well described by the shift current conductivities only when considering Hartree interactions. Indeed, while the peaked response at CNP (*L1*) is captured by both the interacting and non-interacting $\sigma_{xxx}$, the sharp sign-changing response (*L2*) is unique to the Hartree $\sigma_{xxy}$ (Fig. 3e). This distinction is crucial, since it provides the exact ingredients to observe the polarization phase drifts close to CNP. This is seen in Fig. 3d/e (right), which plot simulated photocurrent maps showing the polarization phase drift is unique to the interacting case.

Hartree interactions strongly modify the $\sigma_{xxy}$ components while leaving $\sigma_{xxx}$ largely unaltered. This difference stems from the microscopic components of shift current, where different states in the moiré BZ contribute differently. Specifically, Hartree interactions strongly affects states near the K/K' points of the moiré BZ[34], (Fig. 3f); the regions that contribute to the $\sigma_{xxy}$ component (See Supplementary Section 6). The polarization phase drift through the CNP is attributed to the two distinct filling-dependent conductivities and hence serves as a photocurrent fingerprint of Hartree interactions in TBG. Moreover, the significant modifications around the CNP demonstrate that even minimal doping significantly distorts the quantum geometry of bands.

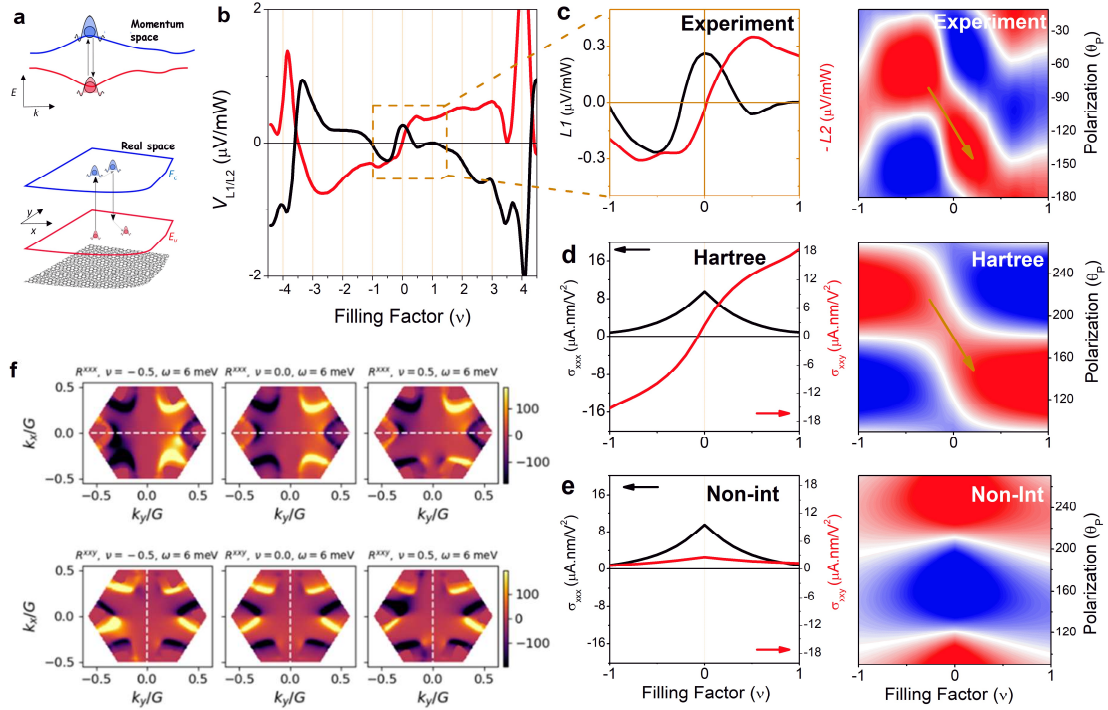

**Figure. 3 Hartree Interaction induced terahertz photocurrents**. **a,** Schematic illustrating shift currents. Top Panel: illustration of interband transitions. Bottom panel: the real-space corresponding to the valence (red) and conduction band (blue) manifold. Electrons experiences real-space shifts of the wave packet when transitioning between bands. **b,** experimentally extracted *L1* and *L2* as a function of filling factor (ν) in D0.94. **c,** left panel: zoom of **b** close to CNP. Right panel: zoom of Fig. 2b close to CNP. **d,e** Left panels: Shift current conductivities as a function of (ν) at 6 meV excitation with (**d**) and without (**e**) Hartree interactions. $\sigma_{xxx}$ on left axis (black) and $\sigma_{xxy}$ on right axis (red). Right panels: simulated photocurrent maps using the calculated $\sigma_{xxx}/\sigma_{xxy}$ with (**d**) and without (**e**) Hartree interactions. Yellow arrows indicate the direction of the polarization phase drift. Qualitatively similar behaviours are observed for different modelling parameters including wavelength, twist-angle and hopping parameters (Supplementary Section 6). (**f**) k-space plots of the shift vector integrand $R_{\alpha\beta c}$ for the two components of the conductivity. Different columns and rows indicate different ν and components, respectively. The dashed white line represents a high symmetry axis for which we plot the case with and without Hartree interactions either side of the hexagon.

**Broken symmetry states**

On the general grounds of symmetry analysis, the presence of shift currents in our devices implies that $C_{2z}T$ symmetry must be broken and a necessary gapping of flat-bands[10] in all our devices. This is at odds with the fact that only D1.03 was explicitly aligned with hBN and exhibited an insulating state at the CNP, whereas D0.94/D1.12 was metallic. These observations are in line with quantum transport experiments where broken $C_{2z}T$ symmetry phases were observed despite the absence of extrinsic symmetry breaking factors[6,23], and point towards an underlying hidden symmetry. In particular, the insulating behaviour observed at the CNP in non-aligned devices has sparked considerable interest in the ground state[23,35,36] where several interaction-driven phases have been predicted. Furthermore, recent tight-binding calculations demonstrated that hBN has a significant impact on TBG's electronic spectrum due to reconstruction effects and strain fields that break the sub-lattice symmetry even for alignment angles larger than $6^{\circ}$[37,38]. Calculations of the bands for D0.94/D1.12 indeed demonstrated that gaps of 3-4 meV emerge, caused by the alignment of hBN with graphene within $4^{\circ}$ (Supplementary Section 9/10). Since we observe polarization-dependent photocurrents for all $\nu$, we believe substrate effects play a major role. This is further supported by the observation of bulk photocurrents in TBG devices at non-magic angles (see Supplementary Section 11), where interactions are less relevant, and measurements between different contact pairs allowed us to map out the C3 symmetry of the system.

To better understand the gapped state at the CNP and integer-filling features, we studied the temperature dependence of the photoresponse. Figure 4a/b plots the photovoltage measured in D1.12/D0.94 for different temperatures. In both cases, the photoresponse decreases rapidly as T approaches 50 K. This is expected for interband response, which is suppressed by temperature smearing of the Fermi-Dirac distribution[19] when approaching the energy scale of the flat bands. However, the CNP response and integer-filling features disappears much faster at 30 K. Although the system still exhibits a polarization dependence, the phase drifts at CNP (Fig. 4c) and integer fillings (Fig. 4d) disappear. Figure. 4e,f further highlights this through plots of $\alpha(\nu)$ for D1.12/D0.94 for two temperatures. The complex temperature dependence in $\alpha$ hence suggests an interplay of symmetry breaking induced by both the hBN substrate and interaction effects.

Our measurements demonstrate that photocurrent is more sensitive to probing gapped states than quantum transport[13]. This is because the latter requires well-defined gapped regions on mesoscopic length scales, which are easily shunted by edges[39] or domains[40]. Bulk photocurrent measurements, however, are not sensitive to un-gapped regions in the same way. The gapped regions generate the photoresponse and the un-gapped regions serve only as current collecting leads. Indeed, our measurements between different contact pairs show $\alpha$ has a non-trivial dependence, likely reflecting spatial variations in local symmetry (Supplementary Section 12). Hence, the gap in our non-aligned devices is likely localized to a specific region[36], which would be undetectable in quantum transport measurements.

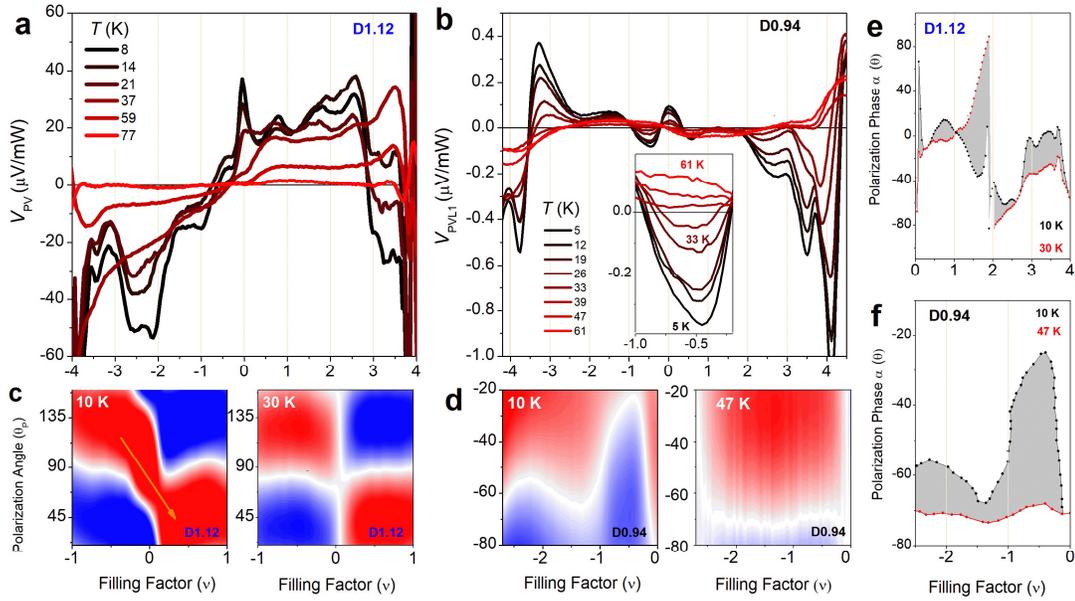

**Figure. 4. Temperature dependence of the photoresponse and polarization phase. a,** Photovoltage $V_{PV}$ (µV/mW) as a function of filling factor (ν) for different temperatures from 8 to 77 K in D1.12. **b,** Same as **a** but for the linear component of the photovoltage $V_{Lin\_L1}$ (µV/mW) measured at different $T$ in D0.94. Inset: zoom around ν = 1. **c,** Polarization dependent component of the photovoltage ($V_{Lin}$) as a function of Polarization angle ($θ_P$) and filling factor (ν) measured in D1.12 for 10 (left panel) and 30 K (right panel) zoomed around CNP. Colour scale: blue to red from -50: 50 µV/mW. **d,** same as **c,** for D0.94 at 10 K (left panel) and 47 K (right panel) zoomed around integer fillings. Colour scale: blue to red from -0.8: 0.8µV/mW. **e,f** polarization phase (α) measured as a function of filling factor (ν) for low and high temperatures in D1.12 (**e**) and D0.94 (**f**). Shaded areas highlight differences in α.

## Cascade phenomena

In addition to the sharp polarization phase drifts at the CNP, we observed additional peaks/dips or drifts in α close to integer fillings, which smear with temperature (Fig. 4). Figure 5b plots $V_{PV}$ measured between different contact pairs in D1.03, the device where signatures of alignment with hBN manifest in quantum transport and near-field measurements performed previously[41]. Interestingly, we find the photoresponse is markedly different. Not only is the peaked response at CNP suppressed, but $V_{PV}$ also exhibits strong oscillations changing sign close to integer fillings. The former can be attributed to strain effects, which are amplified by second-order superlattices[41] (Supplementary Section 6). However, the oscillatory structure cannot be accounted for by a single-particle picture. Notably, with changing polarization, the behaviour around ν = 1 resembles the behaviour at CNP observed in other devices: a peaked response evolving smoothly from a sign-changing response (Fig. 5c). Moreover, the sign-changing response is accompanied by polarization phase drifts through all integer fillings (Fig. 5d-f).

The polarization phase drifts, mirroring that at the CNP (Fig. 3c), reflect the behaviour intrinsic to the cascade of phase transitions[25,26,42] in magic-angle TBG where the chemical potential resets repeatedly at integer flings. In this case, our photocurrent measurements shed light on several empirical observations. First, according to our model (Fig. 3) the polarization phase drifts suggest these states may be gapped like the CNP. Second, the pronounced polarization dependence at ν = 1 compared with other fillings (Fig. 5c) suggests that the symmetry of the underlying many-body ground states may vary across different fillings[43,44]. This variation is attributed to the polarization dependence at ν =

1, which requires two distinct and independent components of second-order conductivity, in contrast to ν = 2 & 3. Third, the sign change in photoresponse at integer fillings reveals that the electronic orders differ post-reset, reflecting subtle changes in the wave functions. Adopting the Stoner-like interpretation of Ref. **26**, where the carriers preferentially occupy one of the four degenerate valleys and/or spin-polarized bands[26], our photocurrent measurements appear to track these phase transitions, with a sign change following each reset. While the shift current is insensitive to valley polarization, the linear injection current is[14,13], suggesting that the oscillatory structure reflects the interplay of flavour-polarization with the spontaneous breaking of time-reversal symmetry. Concurrently, recent low-temperature STM experiments have revealed the presence of inter-valley coherent states at half-filling[36,45]. This observation, combined with our measurements, calls for a more profound exploration of how flat bands and their quantum geometry deform during the cascade, for example, through the heavy fermion model[46,47]. Finally, we note that cascade features could be observed in a fourth TBG sample without commensurate alignment to hBN albeit with contributions from junction effects via the PTE effect (see Supplementary Fig. S8), which seemingly dominates in the case of weaker C2z symmetry breaking.

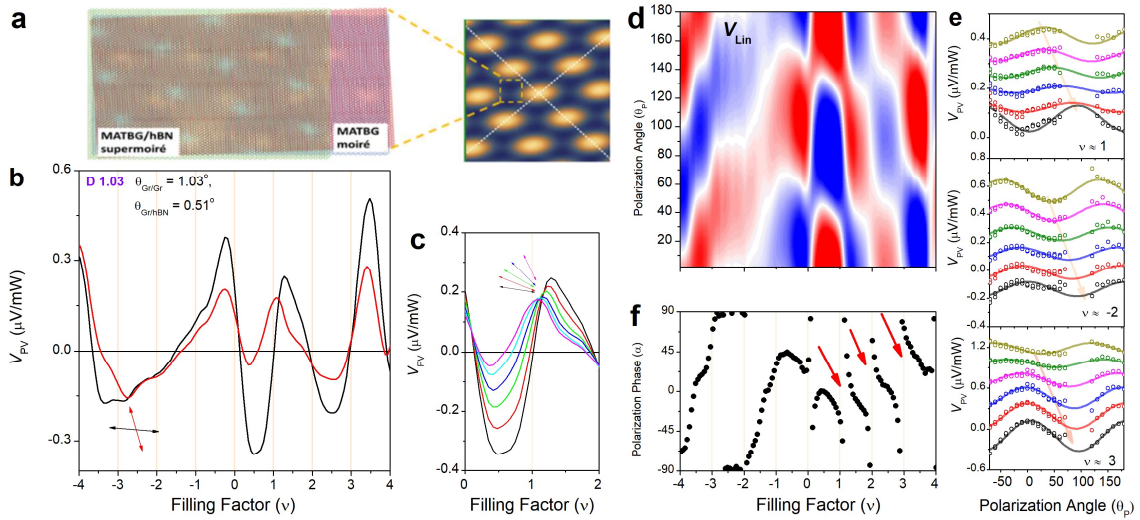

**Figure. 5 Photocurrent cascade in commensurate twisted bilayer graphene aligned to hexagonal-boron nitride with a supermoiré potential**. **a,** Schematic of twisted bilayer graphene aligned to hexagonal boron nitride for commensurate twist-angles where a graphene-graphene and graphene-hBN superlattice bare similar periodicities resulting in a supermoiré structure. Right panel plots a supermoiré lattice in the presence of weak strain (∼ 0.1 %) adapted from Ref. 43. **b,** Photovoltage measurements made between a different contact pair in D 1.03 for θ_p = 10° (black) and 60° (red) respectively. **c,** zoom of **b**, close to ν = 1. **d,** $V_{Lin}$ (θ,ν) for D1.03 with same pair of contacts as in **b**. **e,** $V_{PV}$(θ) plotted for different ν close to integer filling. Open circles are experimental data whilst solid lines are sinusoidal fits. **f,** polarization phase (α) as a function of filling factor plotted in D1.03 The red arrows highlight drifts in α at integer fillings observed in D1.03.

**Discussion**

The occurrence of a terahertz bulk photovoltaic effect in TBG devices without strict alignment to hBN suggests the substrate plays a more subtle role in determining their symmetry and many-body phases. Although further work is required to understand the photoresponse at high $\nu$ and the details of the cascade features, our work demonstrates the sensitivity of photocurrent in probing hidden symmetry of the underlying ground state not easily probed via conventional transport measurements. The rising interest in the quantum metric tensor and its strong influence on electromagnetic response has long called for a technique that is fast and accessible. The direct connection between second-order conductivities and the quantum geometry underlying the terahertz photocurrent response provides a fast and powerful tool in this direction. Additionally, the impact of strain should be further explored because of its known influence on the phase diagram of TBG and strong influence on second-order conductivities (Supplementary Section 6).

From a technological perspective, the measured devices hold strong prospects for terahertz applications. We find extrinsic voltage responsivities around 200 mV/W, which are large considering the device area is 1000 times smaller than the illumination area, and project intrinsic noise equivalent power (NEP) of $10^{-12}$ W/Hz$^{0.5}$ (Supplementary Section 13). These metrics are competitive with previous low-temperature detectors[48] offering major advantages including facile scalability and extraordinary polarization sensitivity. Specifically, moiré materials can be engineered in multilayer configurations to enhance extrinsic absorption while preserving their desirable few-layer properties including a large THz quantum geometric response[49,50]. Moreover, the complex polarization and wavelength dependencies present exciting prospects for on-chip THz polarimetry applications[17].

**Acknowledgements:** R. B acknowledges funding from the European Union's Horizon 2020 research and innovation programme under the Marie Skłodowska-Curie grant agreement No 847517. J.B. acknowledges support from the European Union's Horizon Europe program under grant agreement 101105218. E. K acknowledges Marie-Sklodowska-Curie Fellowship, Project SUPERTERA, Ref. 101033521. P.A.P, Z.Z and F.G acknowledges support from the "Severo Ochoa" Programme for Centres of Excellence in R\&D (Grant No. SEV-2016-0686). Z.Z. acknowledges support funding from the European Union's Horizon 2020 research and innovation programme under the Marie Skłodowska-Curie grant agreement No 101034431 and from the ``Severo Ochoa" Programme for Centres of Excellence in R\&D (CEX2020-001039-S / AEI / 10.13039/501100011033). P.A.P and F.G. acknowledge funding from the European Commission, within the Graphene Flagship, Core 3, grant number 881603 and from grants NMAT2D (Comunidad de Madrid, Spain), SprQuMat and (MAD2D-CM)-MRR MATERIALES AVANZADOS-IMDEA-NC. S.B.P acknowledges the support of the "Presencia de la Agencia Estatal de Investigación" within the "Convocatoria de tramitación anticipada, correspondente al año 2020, de las ayudas para contractos predoctorales (Ref. PRE2020-094404) para la formación de doctores contemplada en el Subprograma Estatal de Fromación del Programa Estatal de Promoción del Talento y su Empleabilidad en I+D+i, en el marco del Plan Estatal de Investigacón Científica y Técnica de Innovación 2017-2020, cofinanciado por el Fondo Social Europeo". E.I. and C.S acknowledge funding from the European Union's Horizon 2020 research and innovation program under grant agreement No. 881603 (Graphene Flagship) and from the European Research Council (ERC) under grant agreement No. 820254, the Deutsche Forschungsgemeinschaft (DFG, German Research Foundation) under Germany's Excellence Strategy - Cluster of Excellence Matter and Light for Quantum Computing (ML4Q) EXC 2004/1 - 390534769. H.A. acknowledge funding from the European Union's Horizon 2020 research and innovation programme under Marie Skłodowska-Curie grant agreement no. 665884. K.W. and T.T. acknowledge support from the JSPS KAKENHI (Grant Numbers 20H00354 and 23H02052) and World Premier International Research Center Initiative (WPI),

MEXT, Japan. J. S acknowledges the Singapore Ministry of Education Tier 2 grant MOE-T2EP50222-0011. G.R. expresses gratitude for the support by the Simons Foundation, and the ARO MURI Grant No. W911NF-16-1-0361 and the Institute of Quantum Information and Matter. C.L. was supported by start-up funds from Florida State University and the National High Magnetic Field Laboratory. The National High Magnetic Field Laboratory is supported by the National Science Foundation through NSF/DMR-2128556 and the State of Florida. PJH acknowledges support by the National Science Foundation (DMR-1809802), the STC Center for Integrated Quantum Materials (NSF Grant No. DMR1231319), the Gordon and Betty Moore Foundation's EPiQS Initiative through Grant GBMF9463, the Ramon Areces Foundation, and the ICFO Distinguished Visiting Professor program. This material is based upon work supported by the Air Force Office of Scientific Research under award number FA8655-23-17047. Any opinions findings, and conclusions or recommendations expressed in this material are those of the authors and do not necessarily reflect the views of the United States Air Force. C. S. acknowledges the supported by the FLAG-ERA grant PhotoTBG, by the Deutsche Forschungsgemeinschaft (DFG, German Research Foundation) - 471733165. R. K K and F. H. K acknowledges funding by MCIN/AEI/ 10.13039/501100011033 and by the "European Union Next GenerationEU/PRTR" PCI2021-122020-2A within the FLAG-ERA grant [PhotoTBG], by ICFO, RWTH Aachen.

**Author Contributions**: R. K K, K. N and F. H. K conceived the experiments. R. K. K and R. B performed photocurrent measurements and analysed the data with support from K. N. G. Li fabricated D0.94, D1.12 & D1.5 and performed quantum transport measurements in these devices. P. S fabricated D1.03. S. C performed calculations and provided theoretical support regarding the shift current. J. M. P fabricated D1.02 supported by P. J. H. S. C performed numerical simulations. Z. Z and P. A. P performed tight-binding calculations of twisted bilayer graphene on hBN with support from F.G. R. K. K, H. A and A. R. P built the cryogenic terahertz photocurrent set-up where measurements were performed. S. B. P and J. B performed supporting photocurrent measurements. M. C provided fabrication support and technical expertise. E. I fabricated bilayer graphene devices supported by C. S. G. P fabricated monolayer graphene devices. E. K provided technical assistance with terahertz measurements. T. T and K. W provided high quality hexagonal-boron nitride crystals. G. R provided theoretical support. J. C. W, C. L and F. H. K supervised the project. R. K. K, R. B, C. L and F. H. K wrote the manuscript with input from all the authors.

**Methods**

**Device fabrication:** Our devices (Fig. S1) were fabricated using standard methods in 2D materials fabrication[51]. For twisted bilayer graphene, the heterostructures were assembled using the "cut and stack" method[52]. In all devices, graphite gates were incorporated into the heterostructures to allow for pristine electrical gating. Following heterostructure assembly, nanolithography was performed on the mesa and gold leads to electrically contact the channel. In D0.94, D1.12 and D1.5 and D1.03 selective etching ($SF_6$) and ($O_2$) was used to expose the graphene channel without etching through the gate. Subsequent deposition of Cr/Au (5/60 nm thick) was made via electron beam and thermal evaporation respectively. This method ensures that the entire active channel is gate-tuneable and removes possible junctions that can arise due to graphite gates (see Supplementary Section 5). In D1.02, and bilayer graphene/monolayer graphene devices, the device was contacted in regions extending outside of graphite gates. Finally, we note that some devices were fabricated on low-doped $Si/SiO_2$ (285 nm) substrates (D0.94, D1.12, D1.02, D1.5 and monolayer graphene) and others on highly-doped $Si/SiO_2$ (285 nm) substrates (bilayer graphene devices, and D1.03). These differences may result in different extrinsic absorptions of the terahertz radiation depending on the choice of substrate.

**Cryogenic photocurrent and electrical measurements:** quantum transport and terahertz photocurrent measurements were performed on an Advanced Research System 4 K cryostat with optical access and temperature control. Electrical measurements were performed using standard AC lock-in techniques (Stanford) in combination with DC voltage sources (Keithley) for electrically gating our devices. Hall effect measurements were performed in the same cryostat using an electromagnet at 0.7 T. Independently, the cryostat could be raised out of the magnet and placed in the path for optical coupling. For terahertz illumination, a gas laser (Edinburgh Instruments) was used and focused into the cryostat using a lens on a cage system with a focal length of 20 mm. To rotate the polarization, we employed polarizers (Purewave) in combination with quarter wave plates (Tydex). The quarter wave plate rotates the incident beam of polarization allowing one to arbitrarily select a polarization direction with the polarizer placed following. In some cases, only a single polarizer was used to rotate the polarization. We verified that the photoresponse was linear with the power ensuring that normalizing by the incident power reflects the polarization dependence of the system. We also checked that performing polarization dependence measurements with a single polarizer, or with a wave plate/polarizer geometry revealed identical results despite the latter enabling larger illumination powers. Experimental data was taken by fixing the polarization dependence and sweeping the gate voltage. Before and after each measurement the power was measured using a room temperature bolometer coupled via a flip mirror placed in the optical path.

**Photodetector performance estimates:** Because of the long-wavelength intrinsic to terahertz radiation, our devices are significantly smaller than the focused terahertz beam. Consequently, a significant portion of the incident radiation is not absorbed. Hence, normalizing the photovoltage by the measured power at the room temperature bolometer (see methods) is not representative of the inherent responsivity of our devices. To estimate the capabilities of our TBG devices for terahertz detection, we project the possible noise equivalent powers (NEP) that may be obtainable in optimized devices with antenna integration. For this, we estimate the total power incident on the sample by calculating the fraction of the incident terahertz beam that is occupied by the device. Considering the wavelength at 0.7 THz = 400 $\mu$m, a numerical aperture (NA) of 0.12, the diffraction limited beam spot diameter is estimated to be 1.8 mm. Given our devices are at most 10 x 30 $\mu$m in size, we find the device area (3E-10 m$^2$) around 8000 times smaller than the area of the focused gaussian beam (2.5E-6 m$^2$). Hence, the power axis on Fig. S16 is normalized by this factor 8000 to estimate the potential responsivities and NEPs obtainable for example in a device with integrated antennas or if the device active area was up-scaled. Because the device operates under zero-bias voltage, we attribute the intrinsic noise floor to Johnson noise , with $R$ = 10,000 $\Omega$ (two-probe resistance of our devices), $T$ = 10 K and  = 1 Hz. Extrapolating from Fig. S16 we find NEP = 7E-13 W/Hz$^{0.5}$. The power dependent measurements in Fig. S16 was performed for $\theta_P$ = 50$^\circ$ illumination.

# References

1. Cao, Y. *et al.* Unconventional superconductivity in magic-angle graphene superlattices. *Nature* **556**, 43–50 (2018).

2. Cao, Y. *et al.* Correlated insulator behaviour at half-filling in magic-angle graphene superlattices. *Nature* **556**, 80–84 (2018).

3. Andrei, E. Y. & MacDonald, A. H. Graphene bilayers with a twist. *Nat. Mater.* **19**, 1265–1275 (2020).

4. Sharpe, A. L. *et al.* Emergent ferromagnetism near three-quarters filling in twisted bilayer graphene. *Science.* **365**, 605-608 (2019).

5. Serlin, M. *et al.* Intrinsic quantized anomalous Hall effect in a moiré heterostructure. *Science.* **367**, 900–903 (2020).

6. Stepanov, P. *et al.* Competing Zero-Field Chern Insulators in Superconducting Twisted Bilayer Graphene. *Phys. Rev. Lett.* **127**, 197701 (2021).

7. Andrei, E. Y. *et al.* The marvels of moiré materials. *Nat. Rev. Mater.* **6**, 201–206 (2021).

8. Tschirhart, C. L. *et al.* Imaging orbital ferromagnetism in a moiré Chern insulator. *Science.* **372**, 1323–1327 (2021).

9. Grover, S. *et al.* Chern mosaic and Berry-curvature magnetism in magic-angle graphene. *Nat. Phys.* **18**, 885–892 (2022).

10. Cea, T., Pantaleón, P. A. & Guinea, F. Band structure of twisted bilayer graphene on hexagonal boron nitride. *Phys. Rev. B* **102**, 155136 (2020).

11. Kazmierczak, N. P. *et al.* Strain fields in twisted bilayer graphene. *Nat. Mater.* **20**, 956–963 (2021).

12. Carr, S., Fang, S., Zhu, Z. & Kaxiras, E. Exact continuum model for low-energy electronic states of twisted bilayer graphene. *Phys. Rev. Res.* **1**, 13001 (2019).

13. Ma, Q., Krishna Kumar, R., Xu, S.-Y., Koppens, F. H. L. & Song, J. C. W. Photocurrent as a multiphysics diagnostic of quantum materials. *Nat. Rev. Phys.* **5**, 170–184 (2023).

14. Ma, Q., Grushin, A. G. & Burch, K. S. Topology and geometry under the nonlinear electromagnetic spotlight. *Nat. Mater.* **20**, 1601–1614 (2021).

15. Sipe, J. E. & Shkrebtii, A. I. Second-order optical response in semiconductors. *Phys. Rev. B* **61**, 5337–5352 (2000).

16. Nakamura, M. *et al.* Shift current photovoltaic effect in a ferroelectric charge-transfer complex. *Nat. Commun.* **8**, 281 (2017).

17. Ma, C. *et al.* Intelligent infrared sensing enabled by tunable moiré quantum geometry. *Nature* **604**, 266–272 (2022).

18. Duan, J. *et al.* Giant Second-Order Nonlinear Hall Effect in Twisted Bilayer Graphene. *Phys. Rev. Lett.* **129**, 186801 (2022).

19. Chaudhary, S., Lewandowski, C. & Refael, G. Shift-current response as a probe of quantum geometry and electron-electron interactions in twisted bilayer graphene.

20. Kaplan, D., Holder, T. & Yan, B. Twisted photovoltaics at terahertz frequencies from

momentum shift current. *Phys. Rev. Res.* **4**, (2022).

21. Arora, A., Kong, J. F. & Song, J. C. W. Strain-induced large injection current in twisted bilayer graphene. *Phys. Rev. B* **104**, L241404 (2021).

22. Pantaleón, P. A., Low, T. & Guinea, F. Tunable large Berry dipole in strained twisted bilayer graphene. *Phys. Rev. B* **103**, 205403 (2021).

23. Lu, X. *et al.* Superconductors, orbital magnets and correlated states in magic-angle bilayer graphene. *Nature* **574**, 653–657 (2019).

24. Xie, Y. *et al.* Spectroscopic signatures of many-body correlations in magic-angle twisted bilayer graphene. *Nature* **572**, 101–105 (2019).

25. Wong, D. *et al.* Cascade of electronic transitions in magic-angle twisted bilayer graphene. *Nature* **582**, 198–202 (2020).

26. Zondiner, U. *et al.* Cascade of phase transitions and Dirac revivals in magic-angle graphene. *Nature* **582**, 203–208 (2020).

27. Xu, X., Gabor, N. M., Alden, J. S., van der Zande, A. M. & McEuen, P. L. Photo-Thermoelectric Effect at a Graphene Interface Junction. *Nano Lett.* **10**, 562–566 (2010).

28. Candussio, S. *et al.* Edge photocurrent driven by terahertz electric field in bilayer graphene. *Phys. Rev. B* **102**, 45406 (2020).

29. Castilla, S. *et al.* Fast and Sensitive Terahertz Detection Using an Antenna-Integrated Graphene pn Junction. *Nano Lett.* **19**, 2765–2773 (2019).

30. Valentin A. Semkin, A. V. S. *et al.* Zero-bias photodetection in 2d materials via geometric design of contacts. *arXiv:2303.16782v1*.

31. Sinha, S. *et al.* Berry curvature dipole senses topological transition in a moiré superlattice. *Nat. Phys.* **18**, 765–770 (2022).

32. He, P. *et al.* Graphene moiré superlattices with giant quantum nonlinearity of chiral Bloch electrons. *Nat. Nanotechnol.* **17**, 378–383 (2022).

33. Chaudhary, S., Endres, M. & Refael, G. Berry electrodynamics: Anomalous drift and pumping from a time-dependent Berry connection. *Phys. Rev. B* **98**, 1–12 (2018).

34. Guinea, F. & Walet, N. R. Electrostatic effects, band distortions, and superconductivity in twisted graphene bilayers. *Proc. Natl. Acad. Sci.* **115**, 13174–13179 (2018).

35. Jiang, Y. *et al.* Charge order and broken rotational symmetry in magic-angle twisted bilayer graphene. *Nature* **573**, 91–95 (2019).

36. Nuckolls, K. P. *et al.* Quantum textures of the many-body wavefunctions in magic-angle graphene. *Nature* **620**, 525–532 (2023).

37. Long, M. *et al.* An atomistic approach for the structural and electronic properties of twisted bilayer graphene-boron nitride heterostructures. *npj Comput. Mater.* **8**, 73 (2022).

38. Song, J. C. W., Shytov, A. V & Levitov, L. S. Electron Interactions and Gap Opening in Graphene Superlattices. *Phys. Rev. Lett.* **111**, 266801 (2013).

39. Zhu, M. J. *et al.* Edge currents shunt the insulating bulk in gapped graphene. *Nat. Commun.* **8**, 14552 (2017).

40. Uri, A. *et al.* Mapping the twist-angle disorder and Landau levels in magic-angle graphene.

*Nature* **581**, 47–52 (2020).

41. Niels C.H. Hesp *et al*, Cryogenic nano-imaging of second-order moiré superlattices. *arXiv2302.05487v1*

42. Choi, Y. *et al.* Correlation-driven topological phases in magic-angle twisted bilayer graphene. *Nature* **589**, 536–541 (2021).

43. Kwan, Y. H. *et al.* Kekul\'e Spiral Order at All Nonzero Integer Fillings in Twisted Bilayer Graphene. *Phys. Rev. X* **11**, 41063 (2021).

44. Lian, B. *et al.* Twisted bilayer graphene. IV. Exact insulator ground states and phase diagram. *Phys. Rev. B* **103**, 205414 (2021).

45. Kim, H. *et al.* Imaging inter-valley coherent order in magic-angle twisted trilayer graphene. *Nature* **623**, 942–948 (2023).

46. Song, Z.-D. & Bernevig, B. A. Magic-Angle Twisted Bilayer Graphene as a Topological Heavy Fermion Problem. *Phys. Rev. Lett.* **129**, 47601 (2022).

47. Datta, A., Calderón, M. J., Camjayi, A. & Bascones, E. Heavy quasiparticles and cascades without symmetry breaking in twisted bilayer graphene. *Nat. Commun.* **14**, 5036 (2023).

48. Bandurin, D. A. *et al.* Resonant terahertz detection using graphene plasmons. *Nat. Commun.* **9**, 4–11 (2018).

49. Khalaf, E., Kruchkov, A. J., Tarnopolsky, G. & Vishwanath, A. Magic Angle Hierarchy in Twisted Graphene Multilayers. **02138**, 1–9 (2019).

50. Park, J. M. *et al.* Robust superconductivity in magic-angle multilayer graphene family. *Nat. Mater.* **21**, 877–883 (2022).

**Supplementary Information**

**S1 - Comparison with the photo thermoelectric effect**

In graphene-based heterostructures, the long lifetime and cooling time of photoexcited hot electrons ensure that its photoresponse exhibits a strong photo thermoelectric (PTE) effect which is driven by the Seebeck coefficients of the materials and temperature gradients induced by heating of the electron gas via photoexcitation[1,2]. Given the large Seebeck coefficients predicted in magic-angle twisted bilayer graphene[3] and a dominant contribution to the terahertz response observed in graphene devices[4,5], one may expect a manifestation of the PTE effect. We explore this possibility by comparing the photoresponse measured in our devices with the functional form expected for the Seebeck coefficient using the semi-classical Mott formula[6] (S1)

$$S\left(E_F\right) = \frac{\pi^2 k_B T}{3e} \frac{1}{G} \frac{dG}{dV_g} \frac{dV_g}{dE_F},$$ (S1)

where $G$ is the 2-probe conductance measured between the same pair of contacts used in photovoltage measurements, and $V_G$ is the gate voltage. Using the density of states (DOS) obtained from the calculated band structures (Fig. 1a main text inset), we obtain the filling-dependent Seebeck coefficient S($\nu$). It is plotted in Fig. S2b for different temperatures calculated from the 2-probe resistance in D0.94 , using the same pair of contacts as photoresponse measurements. In general, the leading contribution to the Mott formula is the transconductance term which is calculated from the gate dependence of the resistivity and results in the oscillatory structure imposed by the integer filling features observable in quantum transport. Figure S2a plots the photoresponse measured in D0.94; we plot the polarization-dependent component of the response $V_{Lin\_L1}$ obtained by taking $V_{Lin\_L1} = V_{PV}$ $_{\theta p = 135} - V_{PV\,\theta p = 45}$ . One can see that the two responses are rather different. The photoresponse exhibits a complex form, peaked at zero doping and increasing gradually approaching full filling $\nu$ = 4. In the PTE contribution, which would follow the estimated Seebeck coefficient, strong oscillations occur likely originating from the changes in transconductance close to integer fillings. This comparison along with the complex polarization dependence (Fig. 2 of main text) provides strong evidence for a contrasting mechanism to the PTE effect.

**S2 – Polarization Dependence in D1.12 and D1.03 samples**

Figure S3 plots the full polarization and filling factor ($\nu$) dependence for two other magic-angle samples presented in the main text, D1.12 and D1.03. The full polarization dependence presented in Fig. 1g,h is plotted in Fig. S3a,d. In both samples the unique behaviour at CNP is observed, that is a peak response at CNP for one polarization direction that evolves smoothly from a sign-changing response. In addition, we find complex polarization dependence can be observed at higher doping levels, especially close to full filling. Like D0.94 (Fig. 2a), the polarization phase ($\alpha$) changes when tuning the filling factor. This is demonstrated in the colour maps which plot the polarization-dependent component $V_{Lin}$ and the white lines that track the changes in $\alpha$, as shown in Fig. 2 of the main text. The raw data are plotted in Fig. S3c/f tracking the changes in $\alpha$ as the filling is changed. Although both dependencies are rather different, they demonstrate a phase drift through CNP (shown for D1.03 in Fig. S3). Some similarities between the two include negative pockets of resistance between integer fillings appearing with polarization and sharp slips in $\alpha$ close to $\nu$ = 4. In summary, Fig. S3 demonstrates that complex polarization-dependent photocurrents manifest profoundly at THz frequencies in magic-angle twisted bilayer graphene.

## S3 – Wavelength dependence of photoresponse in D1.12

As mentioned in the main text, most measurements presented in D1.12 are presented for 0.7 THz (2.7 meV) excitation wavelength rather than 2.5 THz (10 meV). This is because, 0.7 THz showed a much larger response (Fig. 1 of main text) and clearer signatures of the second-order photocurrents including the complex polarization dependence close to the charge neutrality (Fig. 1a of the main text) and drifts in the polarization phase (Fig. S3). The differences are shown in Fig. S4a which compares the photovoltage measured with the same pair of contacts for two different wavelengths 2.5 and 0.69 THz for a fixed polarization direction. Not only is the photoresponse an order of magnitude larger at 0.69 THz, but the dependence on the filling factor is different. The peaked response at CNP is significantly more pronounced and additional features can be discerned, including oscillatory structures at integer fillings and negative/positive pockets of photoresponse between $\nu = 3-4$. The difference in polarization dependence can be seen by comparing the colour maps in Fig. S4b and Fig. S3b, which plots the polarization-dependent component of the photoresponse $V_{Lin}$ for 2.5 THz respectively. Indeed, there is clearly some polarization dependence, but the polarization phase $\alpha$ is significantly more rigid, varying little compared to the case of 0.69 THz. The differences can be seen clearer in Fig. S4c which compares the polarization phase for the two wavelengths and same pair of contacts. In 0.69 THz, the polarization drifts smoothly over 90° from $\nu = 0-4$ whilst in 2.5 THz it is nearly constant with some minor variations. Furthermore, the peak-like structures and drifts at integer filling are clearly only discerned at 0.69 THz (Fig. S4c). The complex wavelength dependence provides further evidence for the origin in second-order response which is very sensitive to wavelength[7,8,9]. Moreover, the significantly larger response suggests we approach the intrinsic THz resonances of the flat bands in magic-angle TBG (Fig. S16) and begin to dominate over any extrinsic polarization axes that may be overwhelming the response at 2.5 THz. Finally, we note that the differences in wavelength may also be due to competing mechanisms, for example, the rigid behaviour of the polarization phase mimics something expected from an antenna (See Supplementary Section 5), while moving to shorter wavelengths we approach the intrinsic bulk response of the system with two principal axes.

## S4 – additional checks of the polarization dependence

Polarization-resolved measurements can be challenging with many artefacts arising. One of the artefacts is that of beam precession, where the photoresponse oscillates or changes owing to a change in spatial position induced by rotating optical elements (waveplates/polarizers) rather than originating from a change in polarization itself. In general, such effects should be minor in our experiment because the focused THz beam is much larger than our device size. Nonetheless, we performed additional checks to verify that the beam precession is minimal and does not significantly influence the measured response. For this purpose, we performed a full 360° degree polarization dependence of the photocurrent using our polarizer. If our measurement is a true polarization effect, then the curves should be identical for polarization directions separated by 180°. Figure S5a-c plots photocurrent measurements performed for different polarization directions and their 180° rotations in D1.12 at 0.7 THz. The solid and dotted lines for each polarization direction represent data with the same polarization direction, but with the polarizer rotated by 180°. Notably, in all cases, the polarization dependences exhibits almost identical photoresponses, except for minute changes. These results demonstrate that our measurements track a change in polarization state whilst beam precession is at a minimum, and justifies the analysis of the polarization dependence within a limited range of $\theta_p$ presented for specific measurements in the manuscript.

**S5 –polarization dependence in non-centrosymmetric 2D materials, near-field enhancements, and junction contributions**

A polarization-dependent photoresponse is typically observed in many different types of mesoscopic devices even in systems that pertain inversion symmetry. Such effects usually originate from extrinsic symmetry breaking of the device architecture such as device edges[10], or near-field enhancements at metallic leads[11]. In asymmetric structures, polarization-dependent photocurrents can be induced that even changes sign with polarization direction[12,13]. Their origin lies in the near-field enhancement of optical fields at the edges of gold contacts. Such near-field enhancements are also polarization-sensitive via the so-called lighting rod effect and can produce a polarization dependent response. To understand possible contributions that could enter our measurements of TBG samples, we measured two control devices. The first, consisted of a single layer graphene (SLG) Hall bar device, with a mesa-like structure similar to that of our TBG devices. The second, is a bilayer graphene (BLG) device that is rather inhomogeneous owing to bubbles and contamination within the heterostructure and contains multiple PN junctions. The motivation for studying such samples is to understand whether inhomogeneity and/or asymmetric device mesas could cause behaviour mimicking the bulk photovoltaic effect.

**Bilayer graphene – Inhomogeneous devices and photoactive junctions far from gold contacts**

First, we present data measured on the BLG device. An optical image is presented in Fig. S6a. It consists of a graphite/hBN/bilayer graphene/hBN/graphite heterostructure. The two graphite contacts are marked in red and dark blue respectively and are used as top and bottom gates to gate the bilayer graphene channel. We note that heterostructure is rather inhomogeneous, with many bubbles in the active area and multiple PN junctions formed near the edge of graphite gates and close to gold contacts. The goal is to understand whether devices with many junctions could cause a non-trivial response. Figure S6c shows the photocurrent as a function of the bottom gate voltage $V_{Graphite}$ for two different polarization directions that are orthogonal to each other. In general, both data have a similar gate dependence, featuring peaks close to zero doping and a function that resembles the Seebeck coefficient in graphene heterostructures, likely originating from the PTE effect[14,2]. The effect of polarization seems to cause only a constant offset in the measured photoresponse. By fixing the doping and varying the polarization angle (Fig. S6b inset) we observe a clear sinusoidal dependence indicative of a polarization dependent response in the system. However, measuring for different doping levels we find the axis of the polarization dependence is fixed. This is clearer seen in Fig. S6e which plots the polarization dependent component of the photocurrent ($V_{Lin}$) as a function of gate voltage and polarization angle, similar to the maps presented in Fig. 2b, Fig. 5d & Fig. S3b,e. It shows that the polarization phase $\alpha$ remains fixed for all the doping levels contrasting greatly with data measured in TBG. Figure 2f plots the extracted coefficients from which we find L1 and L2 are rather different from the offset, D, and are almost entirely independent on the carrier doping. We interpret this behaviour in the following way. In our BG device, there are four junctions. Two of these are formed by the edge of graphite gates (J1 in Fig. S6a), and the other two by gold-BG junctions (J2 in Fig. S6a). Since graphite gates only tune J1, the gate-tuneable photoresponse we measure is likely dominated by those regions. Since the photo-thermoelectric effect is polarization insensitive, it makes sense that the polarization dependent components L1, L2 do not follow its gate dependence. On the other hand, the other two Au-BG junctions are not gate-tuneable. They can however exhibit some polarization dependence via the lightning rod effect, where near fields in the Au are enhanced for certain polarization direction, enhancing the absorption in graphene, and resulting in the coefficients L1, L2 that are independent on the graphite gate. This result indicates that the photo-thermoelectric effect – a mechanism known to dominate in graphene and graphene bilayers – is polarization independent, and the only polarization dependence enters through near-field absorption enhancements close to gold contacts.

**Single Layer Graphene – High quality device with photoactive junctions close to near-field enhancements.**

To further test our understanding of near-field enhancements, we studied a single layer graphene encapsulated with hexagonal-boron nitride device where the gold leads are close to graphite gates and can influence the photoresponse via near-field enhancements (Fig. S7a). The two-probe resistance measured as a function of gate voltage between contact pairs indicated in Fig. S7a is plotted in Fig. S7b revealing a single peaked response close to zero gate voltage when the Fermi-level is tuned to the Dirac point. Figure S7c plots the photovoltage ($V_{PV}$) as a function of the gate voltage for three different polarization directions. For 170° the response is maximum, and its gate dependence mimics what is expected from the photothermoelectric effect changing sign at the Dirac point. Rotating the polarization response lowers the signal smoothly and exhibits a 180° periodicity expected for a polarization-dependent response (Inset of Fig. S7c). This behaviour mimics an antenna/lightning rod effect in-which the response is maximum for one direction and close to zero for the perpendicular when the near-field enhancements are supressed. Indeed, the polarization dependence highlights this approaching zero for 100°. We note however that it also can change sign with the polarization (black curve in Fig. S7c). Nonetheless, tracking the polarization phase through plots of $V_{LIN}(\theta, V_G)$ (Fig. S7d) shows that the polarization phase ($\alpha$) is fixed for all the gate voltages confirming its origin in extrinsic absorption enhancements. Its origin in an antenna effect can be verified by fitting the polarization dependence completely with the function $V_{PV} = L1Sin^2\theta + D$ (red line in the inset of Fig. S7c). The $sin^2$ term captures the polarization dependence governed by an antenna, whilst the offset $D$ captures a polarization independent response in the system (plotted in Fig. S7e). With this, the sign changing response with polarization can be understood as an interplay of an antenna enhanced junction with a polarization independent off-set of opposite sign. When the antenna response is small ($\theta_p = 130°$) the off-set (D) starts to dominate the measurement resulting in a sign-change with polarization. Similar behaviour was observed in graphene PN junction devices previously[11]. In summary, polarization-dependent responses can be expected in any mesoscopic graphene device that preserves inversion symmetry. The key distinction with the second-order response reported in the main text is that the electron system can rotate the principal axes/polarization phase (Fig. 2d), whilst in these devices it is fixed for all the doping levels (Fig. S7d).

**D1.02 Magic-angle twisted bilayer graphene – contributions to the photoresponse by near-field enhancements and isolating the second-order response.**

In this section, we present measurements and analysis of a fourth magic angle device D1.02 for which oscillations in the photoresponse could be observed baring some similarities with D1.03 (Fig. 5 of main text). The device is presented in Fig. S8a. In contrast to the other TBG devices, where the graphite gates are under the gold leads, an un-gated region of TBG exists at the gold-leads in D1.02. The junction, which typically always induces a photovoltage with a polarization dependence caused by near-field enhancements (as in the case of Fig. S6/7), is not gate-tuneable and manifests in our data as a polarization dependent background. This is seen clearly in Fig. S8b which plots the photovoltage measured for three directions. Similar to the case of Bilayer graphene (Fig. S6) the polarization offsets the data and follows a 180° periodicity characteristic of a polarization dependence. However, in contrast to the BLG case, we notice that the functional form with carrier doping is also polarization-sensitive. Therefore, we subtract the background to isolate the polarization dependent component of the photovoltage from the gold-un-gated region whose response is not representative of the second-order photocurrents. Its influence on the polarization dependence can be seen in Fig. S8c which plots $V_{Lin}$ (n,$q_P$) and Fig. S8e which plots the polarization phase as a function of filling factor a(n) accordingly. Indeed, despite the oscillatory structure in a, we notice that the polarization phase is rather rigid, oscillating weakly around a constant value = 80° likely dominated by the near-field enhancements. This indicates that additional gate-tuneable contributions to the polarization phase exist, but are masked by the principal axes of near-field enhancements. According to our control measurements in BLG, this polarization-dependent background contributes to a gate-independent photovoltage offset (Fig. S6c). Therefore, we subtract linear fits to data of Fig. S8b (dashed lines) to isolate the gate-tuneable component of a. The corresponding $V_{Lin}$ ($\nu$,$\theta_P$) maps are plotted in Fig. S8d and resemble now something closer to measurements performed in D1.03 which explicitly has global alignment between TBG and hBN. Specifically, sharp phase changes in a can be seen at various doping levels (Fig. S8f). $V_{PV}$ $_{- Background\ Subtraction}$ is plotted for different filling factors in the vicinity of these drifts in Fig. S8g to visualize and verify the extraction of polarization phase (Fig. S8f). Although the interpretation of the data in D1.02 is slightly more convoluted than the other samples, these measurements and analysis highlight how multiple mechanisms may be contributing to the polarization phase in the THz regime of TBG. The seemingly less prominent second-order response in D1.02 may be due to its lack of strong global alignment to hBN and hence a weaker influence of the symmetry-breaking substrate essential for driving the second-order response.

**S6 – Theoretical analysis of and shift current calculations**

In this section of the Supplemental Materials, we provide the details of the theoretical analysis described in the main text. Specifically, we compute the shift current in twisted bilayer graphene by following Ref. 7 and the references discussed therein. The analysis considers two twist angles, 0.94° and 1.12° where the band structure is calculated using the BM model[15]. Our analysis specifically considered flat-to-flat band transitions only, as flat-to-dispersive transitions are supressed given the low experimental frequency. For a detailed discussion of flat-to-flat and flat-to-dispersive transitions, we refer the reader to Ref. 7.

**6a: Relating experimental and theoretical quantities**

We begin by detailing how microscopically computed shift current can be related to the experimentally relevant quantity photovoltage. Shift current is an example of a non-linear response of an electronic system to an external electric field. We can relate the applied electric field to the measured photocurrent $j$ in general as

$$\begin{pmatrix} j_x \\ j_y \end{pmatrix} = \begin{pmatrix} \sigma_{xxx} & \sigma_{xxy} & \sigma_{xyx} & \sigma_{xyy} \\ \sigma_{yxx} & \sigma_{yxy} & \sigma_{yyx} & \sigma_{yyy} \end{pmatrix} \begin{pmatrix} E_x^2 \\ E_x E_y \\ E_y E_x \\ E_y^2 \end{pmatrix}, \qquad (S2)$$

where we suppressed time and frequency dependence of the electric field. To relate it to the experiment it is helpful to express the polarization of electric field in the polar coordinate system of the crystallographic axis of the material, i.e. $E_x = E \cos \phi$ and $E_y = E \sin \phi$ with $E$ denoting electric field magnitude. In such polar coordinates we have:

$$j_x = \left( \sigma_{xxx} \cos^2 \phi + (\sigma_{xxy} + \sigma_{xyx}) \cos \phi \sin \phi + \sigma_{xyy} \sin^2 \phi \right) E^2 \qquad (S3)$$

$$j_y = \left( \sigma_{yxx} \cos^2 \phi + (\sigma_{yxy} + \sigma_{yyx}) \cos \phi \sin \phi + \sigma_{yyy} \sin^2 \phi \right) E^2 \qquad (S4)$$

In the above Eqs. $\sigma_{ijk}$ are the second-order conductivity tensor components, whose microscopic origins are explained in what follows. We note that the above relation applies to any non-linear second-order response – specific relation to shift current is only enforced when the coefficients $\sigma_{ijk}$ are computed from microscopic principles.

In an experiment, the coordinate system is likely rotated with respect to the crystallographic axis of the system by an angle $\beta$. We thus can carry a standard similarity transform (rotation) of the crystallographic axis x,y to the experimental axis x', y':

$$\begin{pmatrix} j'_x \\ j'_y \end{pmatrix} = \begin{pmatrix} \cos\beta & \sin\beta \\ -\sin\beta & \cos\beta \end{pmatrix} \begin{pmatrix} j_x \\ j_y \end{pmatrix} \qquad (S5)$$

Lastly, the experiment measures photovoltage, which we take as linearly proportional to the photocurrent, i.e. $V_i \propto j_i$. We use the terms photovoltage and photocurrent interchangeably in what follows. The misalignment of the crystallographic axis and the axis of the experiment is a free parameter, which we vary to optimize agreement with observations, but we keep it fixed for a given device. Specifically, we choose the polarization direction ($\theta_P$) for which the peaked response at CNP is observed to define the principle crystallographic axis $\sigma_{xxx}$ corresponding to $\beta$ = 45°, 90° and 132° for D0.94, D1.12, and D1.03 respectively. We note however that this choice of axis may be perturbed by extrinsic contributions to the polarization phase and hence our assignment of the crystal coordinates may not be accurate. Nonetheless, we can compare qualitatively the orthogonal coefficients L1/L2 that we extract from the polarization dependence with the calculated conductivities.

To arrive at the experimental parametrization of the photocurrent introduced in the main text (which is measured along y' axis)

$$V'_y = L_1 \sin 2\theta + L_2 \cos 2\theta + D \qquad (S6)$$

we project angular harmonics of Eq. 4 above to obtain

$$\begin{aligned} L_1 = &\frac{1}{4}\left(\sigma_{xxx} - \sigma_{xyy} + \sigma_{yxy} + \sigma_{yyx}\right)\cos\beta \\ &+ \frac{1}{4}\left(-\sigma_{xxx} + \sigma_{xy} \ + \sigma_{yxy} + \sigma_{yyx}\right)\cos 3\beta \qquad (S7) \\ &- \frac{1}{2}\left(\sigma_{yxx} - \sigma_{yyy} + \left(\sigma_{xxy} + \sigma_{xyx} + \sigma_{yxx} - \sigma_{yyy}\right)\cos 2\beta\right)\sin\beta \end{aligned}$$

$$\begin{aligned} L_2 = &\frac{1}{4}\left(-\sigma_{xxy} - \sigma_{xyx} + \sigma_{yxx} - \sigma_{yyy}\right)\cos\beta \\ &+ \frac{1}{4}\left(\sigma_{xxy} + \sigma_{xyx} + \sigma_{yxx} - \sigma_{yyy}\right)\cos 3\beta \qquad (S8) \\ &+ \frac{1}{2}\left(\sigma_{yxy} + \sigma_{yyx} + \left(-\sigma_{xxx} + \sigma_{xyy} + \sigma_{yxy} + \sigma_{yyx}\right)\cos 2\beta\right)\sin\beta \end{aligned}$$

$$D = \frac{1}{2}\left(\sigma_{yxx} + \sigma_{yyy}\right)\cos\beta - \frac{1}{2}\left(\sigma_{xxx} + \sigma_{xyy}\right)\sin\beta \qquad (S9)$$

for the three coefficients of Eq. 5. In Eq. 4, $\theta = \phi - \beta$ denotes the electric field polarization with respect to the experimental axis. We highlight here that in addition to the two polarization axis-dependent components, the general analysis of Eq. S8 gives a coefficient, here called D, that is, independent on the direction of polarized light.

## 6b: Symmetry dependence

In cases where the system lacks symmetry, there are six distinct components in the second-order conductivity tensor. However, in the absence of strain, the system exhibits $C_{3z}$ symmetry, which imposes the following constraints on the components of the second-order conductivity.

$$\sigma_{xxy} = -\sigma_{yyy} = \sigma_{yxx} \qquad (S10)$$

$$\sigma_{xxx} = -\sigma_{yyx} = -\sigma_{yxy} \qquad (S11)$$

When $C_{2y}$ symmetry is also present, only one independent component (i.e., $\sigma_{xxy}$) remains nonzero. To obtain nonzero $\sigma_{xxx}$ and $\sigma_{xxx}$ components, we introduced a layer-dependent sublattice offset, $\Delta_1 \neq \Delta_2$. The $\sigma_{xxx}$ component exhibits peaks at two frequencies, which are close to the sublattice offset values $\Delta_1$ and $\Delta_2$. In contrast, the $\sigma_{xxy}$ behaviour is characterized by a single peak in frequency, coinciding with the average gap, as discussed in Ref. 7 . Furthermore, the qualitative features of these plots are not influenced by a change in the sublattice offset $\Delta_1$ and $\Delta_2$, which only shifts the peaks in frequency space, as shown in Fig. S9.1 and Fig. S9.2.

In the remainder of this supplement section, we will focus on the dependence of the individual conductivity coefficients as their behaviour can be more straightforwardly traced back to the microscopic origins. We note for reference the expression for the photocurrent and the corresponding experimental coefficients when the $C_{3z}$ symmetry is present in the system, e.g., the system is unstrained:

$$L_1^{C_{3z}} = -\sigma_{xxx} \cos 3\beta + \sigma_{yyy} \sin 3\beta \qquad (S12)$$

$$L_2^{C_{3z}} = -\sigma_{yyy} \cos 3\beta - \sigma_{xxx} \sin 3\beta \qquad (S13)$$

$$D^{C_{3z}} = 0 \qquad (S14)$$

Specifically, if C3z symmetry is present, the polarization-independent term D vanishes and the two components, L1 and L2, take also a more straightforward form where the relation to the non-linear conductivities of Eq. 9,10 is more explicit. We note however the in our experiments a finite D is always present and share some similarities with L1/L2 we attribute to strain effects (S9).

## 6c: Filling dependence

We now discuss the photocurrent behaviour as a function of filling. For simplicity of the analysis, we focus on the case of $C_{3z}$ symmetry as we found that it qualitatively reproduces experimental trends and could be observed experimentally through polarization dependent measurements on D1.5 (see Supplementary Section 11). The two components $\sigma_{xxx}$, $\sigma_{xxy}$ exhibit remarkably distinct responses to changes in filling.

Specifically, the $\sigma_{xxx}$ component reaches its maximum value at zero filling but gradually decreases with increasing non-zero fillings, as depicted in Fig. S9.1. This behavior can be attributed to a substantial contribution from the region where the gap forms due to a finite sublattice offset. Consequently, its magnitude diminishes when transitions from these regions are Pauli-blocked at non-zero fillings. Additionally, it is worth noting that the sign of the xxx component remains unaffected by the filling value and is solely determined by the sign of the difference between $\Delta_1$ and $\Delta_2$.

Conversely, the $\sigma_{xxy}$ component exhibits a markedly distinct response to variations in filling. Initially, it possesses a small, non-zero value at the charge neutrality point (CNP), which increases with higher filling, its sign depending on the filling value. This dependence on sign becomes noticeable only at considerably large fillings, approximately around, $\nu \approx 2$, as depicted in Fig. S9.2.

The two independent components of shift-current conductivity demonstrate a behavior similar to the L1 and L2 components of photocurrent discussed in the main text. However, it's important to note that the observed trend of sign variation in $\sigma_{xxy}$ with filling is too weak to account for the sharp polarization phase slip observed in the photocurrent at the CNP, as shown in Fig. 3e of the main text. A noteworthy characteristic of the $\sigma_{xxy}$ component is its peak frequency, which displays a subtle dependence on the filling values. As shown in Fig. S9.2, the peak position remains unchanged for filling values of opposite signs and equal magnitudes but starts to increase with greater magnitudes of filling values. Furthermore, unlike $\sigma_{xxx}$, the sign of $\sigma_{xxy}$ remains unaltered when $\Delta_1$ and $\Delta_2$ are interchanged.

### 6d: The Role of interactions

Electron-electron interactions in twisted bilayer graphene can significantly affect the band structure and quantum geometry of electronic bands. Previous work in Ref. 7 has established that these interactions can significantly enhance the shift current conductivity compared to non-interacting band structures. In these calculations, we account for electron-electron interactions by incorporating Hartree corrections as we expect that at temperatures larger than ~ 10K, the role of exchange phenomena is suppressed. The effect of Hartree corrections is particularly pronounced in $\sigma_{xxy}$ at finite fillings, resulting in a sharp sign-changing behavior around the charge neutrality point (CNP), where the sign of $\sigma_{xxy}$ depends on the filling's sign. In contrast, $\sigma_{xxx}$ remains largely unaffected by Hartree corrections, primarily due to the main correction originating around the K/K' points shifting these points upwards in energy compared to the Γ point, but retaining the overall Dirac-like linear dispersion. As a result of this, effectively, the band structure near the Γ point becomes flattened (See Ref. 7 and references therein and the discussion in the main text) compared to the K/K' points. Consequently, $\sigma_{xxx}$ consistently retains a peak feature at the CNP.

These substantial modifications in $\sigma_{xxy}$ offer a plausible explanation for the observed polarization phase drift phenomena in the photocurrent experiment, as illustrated in the right panel of Fig. 3. These changes arise from the interaction-induced modifications to the shift current integrand, which exhibit opposing trends for opposite fillings, as indicated by the green circles in Fig. S9.3.

### 6e: The Role of strain

We also study the role of strain (Fig. S9.4) on different components of second-order conductivity. We apply opposite strain to two layers following the formalism in Ref. [16]. Aside from the overall enhancement in the magnitudes of in conductivities, we notice several additional consequences. In particular, the peaked response characteristic to $\sigma_{xxx}$ is seemingly supressed and attains a sign-changing behaviour similar to $\sigma_{xx}$ . In fact, both conductivities have very similar qualitative behaviour with filling factor, which would lead to a more rigid polarization dependence – a photoresponse that depends weaker in the polarization direction of the incident electric field. Both these findings are consistent with our experimental observations regarding the measurements of D1.03 (Fig. 5 main text). In this sample, the peaked response at CNP is less prominent and the polarization dependence is significantly more rigid. Therefore, we attribute these marked differences to a larger inherent strain in D1.03. The difference with other magic-angle samples we believe lies in the presence of the supermoiré potential which acts as a magnifying glass[17] amplifying strain. Our theory is evidenced by the near-field photocurrent maps which show striped phases at length scales of several 100 nanometres. We note that whilst this qualitative analysis offers some explanation to the marked difference in the polarization dependence in D1.03, the second-order superlattice may be playing an even more pivotal role that has so far not been addressed on any theoretical footing.

**S7 – Quantum transport measurement of the $2\omega$ response in twisted bilayer graphene**

Some of the most prominent intraband contributions to the second-order conductivity in twisted bilayer graphene has been shown to originate from quantum rectification effects, in-which the internal symmetry of the system causes a DC or $2\omega$ response when excited with an AC signal of frequency $\omega$. These effects can be measured using $2\omega$ frequency measurements. In our THz measurements, such effects can be observed in the DC response to THz excitation. In other words, the intraband THz response can be thought of as the non-linear quantum rectification in the high frequency limit and therefore should have a similar response to the $2\omega$ quantum transport experiments in the lower frequency limit. To test this, we performed non-linear quantum transport experiments using the same probes from which we measure photoresponse (Fig. S10). The geometries are indicated in the schematics below the figure. The black data plotted on left axis correspond to the quantum transport experiment with current driven along the longitudinal direction of the Hall bar, whereas the red on the right axis plots the photoresponse measured with the same pair of contacts under photoexcitation with polarization direction aligned in the longitudinal direction of the Hall bar. We find the two responses are quite different. The only finite response of the $2\omega$ response is located close to full filling, whilst in the THz photoresponse it is finite also in the bands. Furthermore, a small sign-changing-like feature can be seen in the $2\omega$ response close to charge neutrality, whilst the photocurrent is peaked. This comparison shows the measured photoresponse is rather different to the $2\omega$ response and originates from a distinct mechanism.

**S8 – Temperature dependent quantum transport measurements characterizing the gaps at full filling**

In the main text, we claim that the terahertz excitations in our system lie within the flat bands and excitations between flat and remote bands are negligible because our excitation energies (3-10 meV) are smaller than the band gaps separating the flat and remote bands. To verify this, we present in Fig. S11 quantum transport measurements on the studied devices and extract the band gaps from temperature dependent measurements of the resistive states at full fulling of the moiré Brillouin zone $\nu = \pm 4$. Figure S11a plots the conductivity ($\sigma_{xx}$) as a function of temperature measured within the insulating states for filling factors $\nu = \pm 4$. We find that the conductivity increases dramatically approaching higher temperatures following an exponential trend expected for activation behaviour. The extracted gaps are plotted in Fig. S11b for the studied samples. In general, the gaps are larger than the maximum excitation wavelengths employed in the main text (10 meV). This is clear for all electron doping. For hole doping, in D0.94 and D1.03 the gaps are significantly supressed. This behaviour may be attributed to the alignment to the underlying hBN substrate which can strongly perturb TBGs electronic spectrum. We note that the small gaps on the hole doping side may result in some contributions to the photoresponse from flat-remote band transitions when the system is doped close to $\nu = \pm 4$. Our experimental data, however, still suggests they may be negligible since the response generally attains e-h asymmetry close to full filling mimicking what is observed for electron doping when the gaps are large and flat-remote band are certainly supressed. Moreover, given the finite band widths $\sim (10 - 20$ meV$)$ expected for magic-angle TBG, those transitions can be safely ignored in our analysis of the shift current contributions close to charge neutrality (Fig. 3 of main text).

## S9 – Alignment angles of twisted bilayer graphene to hBN extracted from optical images

As explained in the main text, manifestation of the bulk photovoltaic effect requires the breaking of inversion symmetry. In twisted bilayer graphene, in-plane components of the conductivity are non-zero only in the case of breaking $C_2T$ symmetry. This is typically achieved via alignment to an underlying hexagonal-boron nitride substrate which breaks the sub-lattice degeneracy of graphene imposing a C3 symmetry on the twisted graphene system. Aside from our photocurrent experiments, signatures of this symmetry breaking are usually observed as gap openings in the flat-band due to the breaking of the sub-lattice symmetry[18] and additional satellite peaks in the resistivity originating from the graphene/hexagonal-boron nitride moiré superlattice[19,20,21]. This is clear in our D1.03 (Fig 1d of main text) whilst our other devices show no signatures in quantum transport measurements of alignment to the hBN substrate. Optical images (Fig. S12) highlight that our device may have alignment angles of approximately 4°, usually far from what is expected to have any significant influence on TBGs electronic spectra. However, recent work has demonstrated these alignment effects can manifest even for these large twist-angles due to complex lattice reconstruction effects leading to the breaking of $C_2T$ symmetry (See Supplementary Section 10).

## S10 – Tight-binding calculations of our twisted bilayer graphene encapsulated with hexagonal-boron nitride devices

Using the alignment angles extracted from the optical images we proceed to calculate the band structure of TBG on hBN using an atomistic approach[22]. We consider a tetralayer structure, where our approach first involves the construction of an unrotated structure, where the bonds of graphene and hBN are parallel. The centre of the supercell has an AAAA stacking, where a carbon site of graphene and nitrogen (N) sites of top and bottom hBN share the same in-plane position $(x, y) = (0, 0)$. Next, we rotate the top layer graphene $(\theta_{TBG})$, the top and bottom layer hBN $(\theta_{TOP}, \theta_{BOT})$ with respect to the bottom layer graphene. $\theta_{hBN}$ is the twist angle between hBN layers. We use positive angles to indicate counterclockwise rotations in our notation. It is important to note that, to keep the periodicity of these structures, the lattice constant of hBN is slightly modified. After constructing a commensurate supercell, we use semi-classical molecular dynamics, which is implemented in LAMMPS[23], to fully (both in-plane and out-of-plane) relax the graphene layers. For intralayer and interlayer interactions between graphene, we use the adaptive intermolecular reactive empirical bond order potential (AIREBO-M)[24] and the interlayer potential (ILP)[25], respectively. We use the same RDKC potential for the C-B and C-N with interlayer interaction of 60% and 200% with respect to the original C-C interaction, respectively[26]. The hBN layers are fixed in a flat configuration to mimic a few layers substrate. We assume that the relaxed structures keep the periodicity of the rigid cases. Once the TB Hamiltonian is constructed, we calculate the electronic properties of the encapsulated TBG/hBN system.

The two middle narrow bands of pristine TBG are degenerate at the K and K' points, which are protected by $C_2T$ symmetry. Here, $C_2$ represents the two-fold rotation operator and T represents the time-reversal operator. However, the presence of an hBN substrate induces a mass term (due to a sublattice polarization) that breaks $C_2T$ symmetry, resulting in the opening of a gap in the Dirac cones. This gap is well-known to exist in samples where the hBN is nearly aligned with its adjacent graphene layer. Previous theoretical studies have suggested that this gap persists even for angles between the hBN and graphene that are far from alignment[26,27]. The lattice structure of the tetralayer system and the degree of alignment for the two samples are shown in Fig. S13a. The corresponding band structures are depicted in Fig. S13b and Fig. S13c, where a breaking of inversion symmetry is evident even if each hBN layer is far from alignment. According to Refs 26,27 the persistence of the mass gap is a combined effect of the large on-site potentials of the boron and nitrogen sites and the large strain fields induced by relaxation in the TBG lattice structure. We found that in both D1.12 and D0.94 samples, assuming the alignment angles extracted from optical images in Fig. S12, the encapsulation induces a small mass gap of about $\sim$ 4–6 mev which is strongly sensitive to the degree of alignment between hBN and its adjacent graphene layer. If the degree of alignment is reduced, this mass gap is enhanced.

The hBN generates an effective superlattice potential acting in the adjacent graphene layer. If graphene/hBN is commensurate, then the potential can be expanded in terms of the graphene reciprocal space basis[28], resulting in a uniform and periodic position-dependent mass gap whose magnitude strongly depends on the degree of alignment between graphene and hBN. In addition, an interaction induced Hartree-Fock gap has been predicted at charge neutrality point[29], by fitting a continuum model to the TB bands in Fig. S13 and by using the procedure described in Ref. 18, we have found that the mass gap is increased by 2-4 meV for both twist angles depending on the value of the dielectric constant. If TBG is incommensurate with hBN, the superlattice potential becomes a non-periodic spatially varying field. For small degrees of alignment, the systems tend to stabilize the commensurate condition[30] and the periodic mass gaps are well defined. For large angles we expect local variations of the sublattice polarization. A similar effect is used to describe position-dependent band topology in TBG encapsulated with hBN[31].

### S11 – Second-order response in D1.5 and measurement of C3 symmetry

In second-order response, the underlying symmetry of the system can be mapped out by measuring its angular dependence, which is typically done using a circular mesa or the so-called sun-flower geometry, that allows measurement between different contact pairs for a fixed driving field. In our photocurrent measurements, we measure Hall bar devices which prevent a full mapping of the symmetry in this way. However, the symmetry may also be measured in Hall bar devices, by fixing the choice of contacts and rotating the polarization direction. In this case, the response always exhibits a 180° periodicity, but the polarization phase changes between the choice of contacts and should maintain a strict mapping reflecting the underlying symmetry of the system. Here we provide evidence for an intrinsic C3 symmetry in a non-magic angle device (D1.5) we attribute to the hBN substrate. Figure S14a plots the resistivity of the device as a function of gate voltage. In contrast to the magic-angle samples, no integer filling features can be observed. It does however exhibit marked insulating states from which we extract gaps of around 50 meV and a twist-angle of 1.5°. The sharp peak at CNP is not activated but rather weakens slowly with increasing temperature characteristic of a zero density of states rather than a band gap. The sample hence exhibits no signature of C3 symmetry breaking via quantum transport. Nonetheless, as in the magic-angle samples, it obtains a complex polarization-dependent response. Figure S14b plots the photoresponse measured between the pair of contacts indicated in the schematic inset referred to as "longitudinal geometry". It shows strong peak-like and sign-changing structures with the polarization of light. The orthogonal direction shows similar response except it is phase shifted by 45° (Fig. S14,c,e). Analysing the coefficients L1 and L2 we find that this phase shift is exactly the value that would be observed in a C3 symmetric system and serves as a photocurrent fingerprint of broken inversion symmetry. This mapping between two orthogonal pairs extends for finite doping close to the CNP (Fig. S14d,f) reaffirming the symmetry relation. Upon close inspection, sharp polarization phase drifts in the vicinity of the CNP can also be observed (Fig. S14g) providing further evidence for its intrinsic second-order nature. We note however that the symmetry condition seemingly breaks down close to full filling and within the remote bands we attribute to extrinsic contributions to the polarization dependence (see Supplementary Section 5). This analysis demonstrates that multiple mechanisms may contribute to the polarization phase and may be perturbing our interpretation of the principal/crystallographic axis. However, the ones that are intrinsic to TBGs quantum geometry may still be distinguished via changes in the polarization phase induced by doping. These measurements hence provides unambiguous evidence of the second-order response and intrinsic C3 symmetry of TBG on hBN. In our magic-angle samples we find no strict symmetry relation can be found between different contact pairs which we attribute to its fragile lattice structure, in which domains of varying alignment and even sub-lattice offsets can exist within a single sample and mixing with extrinsic contributions. Nonetheless, within a single choice of contacts, strong changes in their polarization phase induced by doping not only confirm its origin in the second-order response, but also reveal the quantum geometry induced by interaction effects.

### S12 – Measurements between different contact configurations in magic-angle samples

As demonstrated in the Supplementary Section 11, measurements between different contact pairs provide further evidence for the bulk origin of the effect and allows one to discern the symmetry of the system. In our non-magic angle sample (D1.5), we found that the photoresponse measured between two orthogonal pairs of contacts respects that expected for a C3 symmetric system, in which a 45° phase shift is measured in the polarization phase when changing contact pairs by 90°. In our magic-angle samples, this symmetry relation was not so easy to demonstrate and the behaviour was rather more complex. Figure S15 plots the polarization phase measured between different contact configurations in D1.12; the corresponding measurement geometries are sketched on the right-hand side. G3 and G6 depict two measurement geometries where the current axis is orthogonal to each other as in the case of D1.5 (Supplementary Section 11). The two measurements however are rather different, in which the polarization phase of G6 varies much weaker than G3 and no connection can be made between their polarization dependences. This discrepancy suggests that the measurement G6 contains additional contributions to the polarization dependence that may be occurring locally, for example, near-field enhancements which may be larger because of the significantly larger contacts. With the rest of the working contacts available, we performed similar measurements (G3-G6) for the same wavelength and temperatures and notice several further distinct observations. In all contact configurations, marked changes in the polarization phase can be observed. In G1, G2 and G3 the overall trends in the polarization phase are similar. This demonstrates that the changes in the polarization phase are not unique to a certain choice of contacts and provides ample evidence for its bulk origin. Second, the polarization phase drifts at CNP are distinct to certain regions of the device. This suggests that the CNP response may have some contributions from interaction effects which is truly local, in agreement with our quantum transport measurements that cannot discern a gapped state. On the other hand, the presence of the phase drift at CNP may be unique to local unstrained regions of the device. Third, in one geometry the polarization phase drifts in the opposite direction to the others (G5). Physically, this symmetry condition requires one of the conductivities to change sign spatially. This may be the case in the presence of stacking faults which may reverse the sub-lattice offsets and the sign of $\sigma_{xxx}$ accordingly (see Supplementary Section 6). These complex behaviours suggest the symmetry of magic-angle TBG is far more complex than the simple C3 case as excepted for rigid TBG on hBN. We postulate that strain may be responsible for breaking further symmetries, and, combined with local structural variations may be responsible for the complex contact dependencies observed in this system, whilst extrinsic contributions to the polarization phase may also be perturbing the symmetry analysis.

**S13 – Terahertz photodetection in magic-angle twisted bilayer graphene**

As demonstrated in Supplementary Section 3, we observe that the THz response significantly increases around 0.7 THz, by nearly two orders of magnitude for certain doping levels. This is depicted more clearly in the inset of Fig. S16 which plots the wavelength dependence in D1.12 for $\nu$ = -3.5. The sharp increase differing from a finite frequency Drude response[32] distinguishes its origin in the bulk photovoltaic effect and suggests we approach the interband resonances (inset of Fig. 1a). Moreover, the large response is consistently observed for different polarization directions (see Supplementary Section 2), suggesting that the enhancement is intrinsic to TBG rather than attributable to antenna effects. We find extrinsic voltage responsivities around 200 mV/W for optimal doping, which is large considering the device area is 1000 times smaller than the illumination area and a significant portion of the incident power is not absorbed. To estimate the capabilities of our TBG devices for terahertz detection, we project the possible noise equivalent powers (NEP) that may be obtainable in optimized devices with antenna integration (Fig. S16). For this, we estimate the total power incident on the sample by calculating the fraction of the incident terahertz beam that is occupied by the device. Considering the wavelength at 0.7 THz = 400 $\mu$m, a numerical aperture (NA) of 0.12, the diffraction limited beam spot diameter is estimated to be 1.8 mm. Given our devices are at most 10 x 30 $\mu$m in size, we find the device area (3E-10 m$^2$) around 8000 times smaller than the area of the focused gaussian beam (2.5E-6 m$^2$). Hence, the power axis on Fig. S16 is normalized by this factor 8000 to estimate the potential responsivities and NEPs obtainable for example in a device with integrated antennas or if the device active area was up-scaled. Because the device operates under zero-bias voltage, we attribute the intrinsic noise floor to Johnson noise , with $R$ = 10,000 $\Omega$ (two-probe resistance of our devices), $T$ = 10 K and  = 1 Hz. Extrapolating from Fig. S16 we find intrinsic NEP = 7E-13 W/Hz$^{0.5}$. The power dependent measurements in Fig. S16 was performed for $\theta_P$ = 50$^o$ illumination.

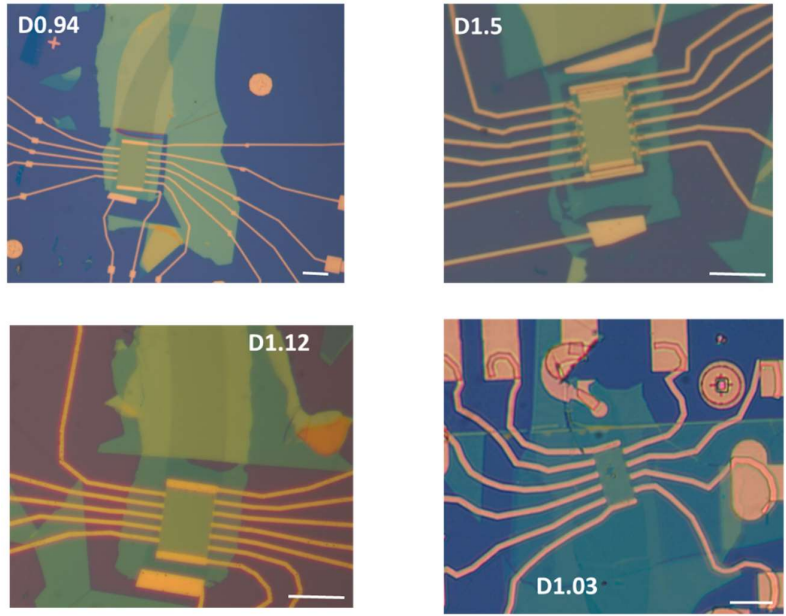

**Figure S1 Optical Images of studied devices.** The four panels show optical images of the studied twisted bilayer graphene devices in this work taken at 100 x or 50 x zoom. Scale bar corresponds to 10 μm.

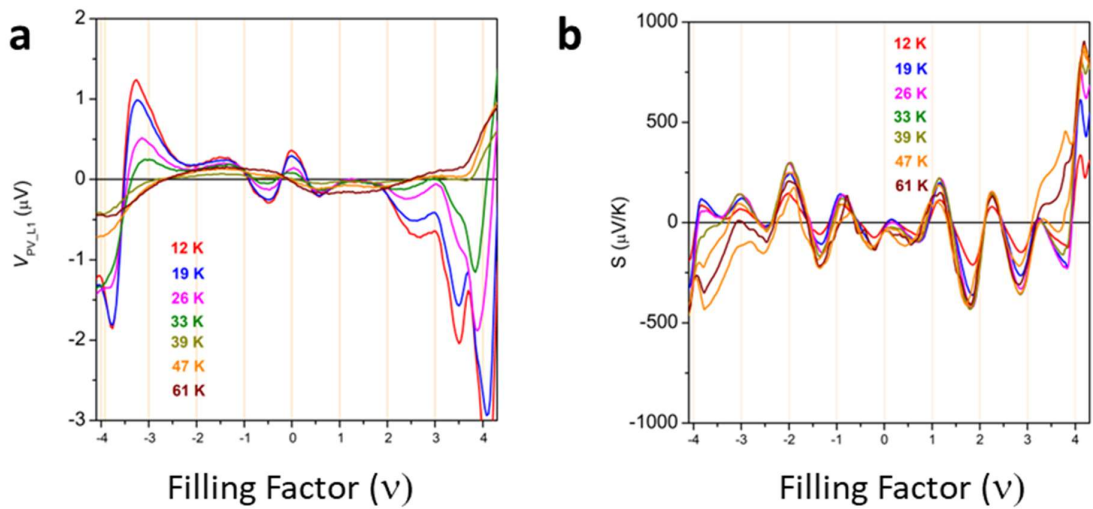

**Figure S2. a,** temperature-dependent measurements of the photovoltage $V_{Lin}$ (μV) as a function of filling factor (ν) measured in D0.94. **b,** Seebeck coefficient (S) calculated via the Mott Formula (S1) plotted as a function of filling factor (ν) for the same temperature range as **a**.

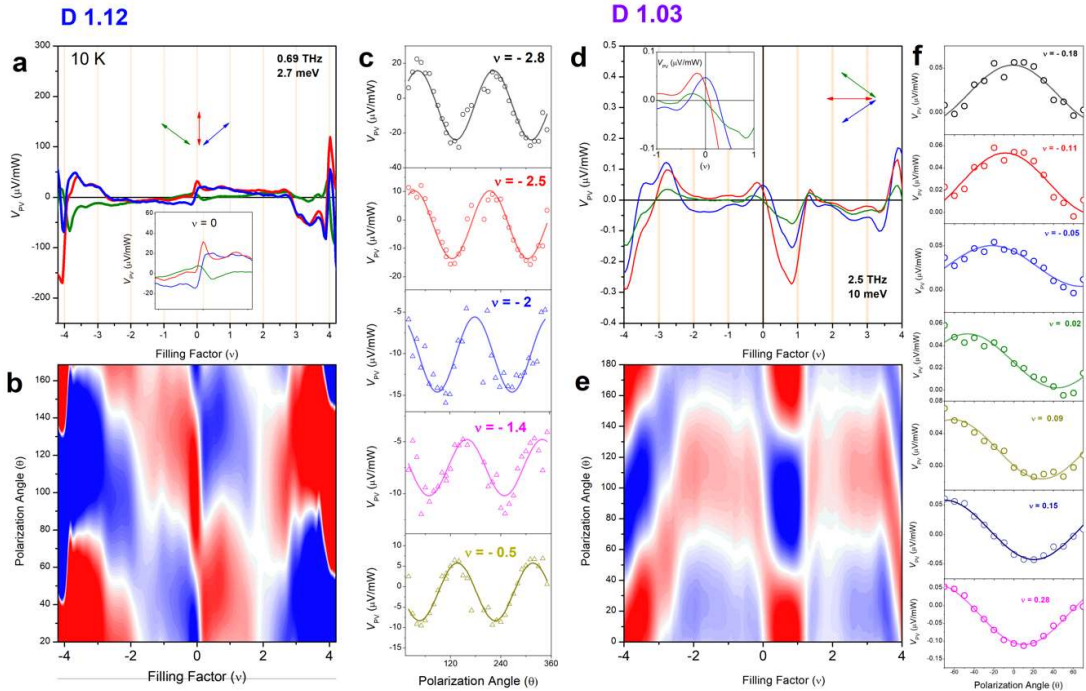

**Figure S3, data for D1.12 (a-c) and D1.03 (d-f). a,** photovoltage ($V_{PV}$) as a function of filling factor ($\nu$) measured in D1.12 for three different polarization directions ($\theta_P$) at 0.7 THz (2.7 meV) excitation. Inset plots zoom around CNP. **b,** polarization dependent component of the photoresponse ($V_{Lin}$) as a function of $\nu$ and $\theta_P$ for the same pair of contacts as in **a. c,** $V_{PV}$ ($\theta_P$) for different doping levels. Open circles are experimental data and solid lines are sinusoidal fits. **d,** same as **a** for D1.03. Polarization directions are chosen to highlight the emergence of the peaked response at CNP. **e,** same as **b** for D1.03. **f,** same as **c** plotted for doping levels in the vicinity of the CNP highlighting the phase drifting behaviour.

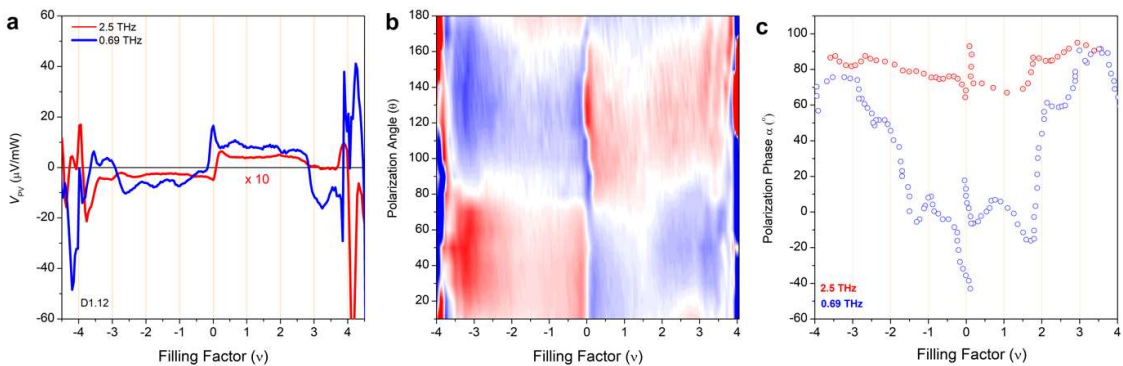

**Figure S4. Wavelength dependence of THz response. a,** Photovoltage ($V_{PV}$) normalized by the incident power for a fixed polarization direction ($\theta_P = 90^\circ$) measured at 0.69 THz (blue) and 2.5 THz (red); 2.5 THz data is multiplied by 10 to compare its functional form with 0.69 THz. **b,** Polarization dependent component of the photovoltage ($V_{Lin}$) as a function of polarization angle ($\theta_P$) and filling factor ($\nu$) measured at 2.5 THz. **c,** Polarization phase extracted from maps Fig. S3b and **b** for 2.5 and 0.69 THz respectively.

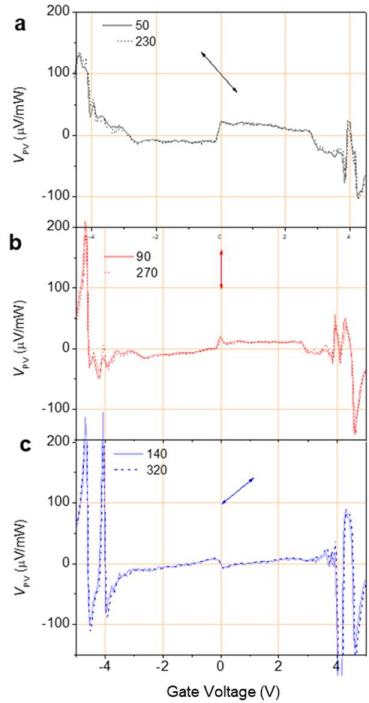

**Figure S5 Checking for artefacts. (a-c)** Photovoltage $V_{PV}$ as a function of gate voltage for fixed polarization directions. Solid and dashed lines plot measurements made for polarization directions 180° apart.

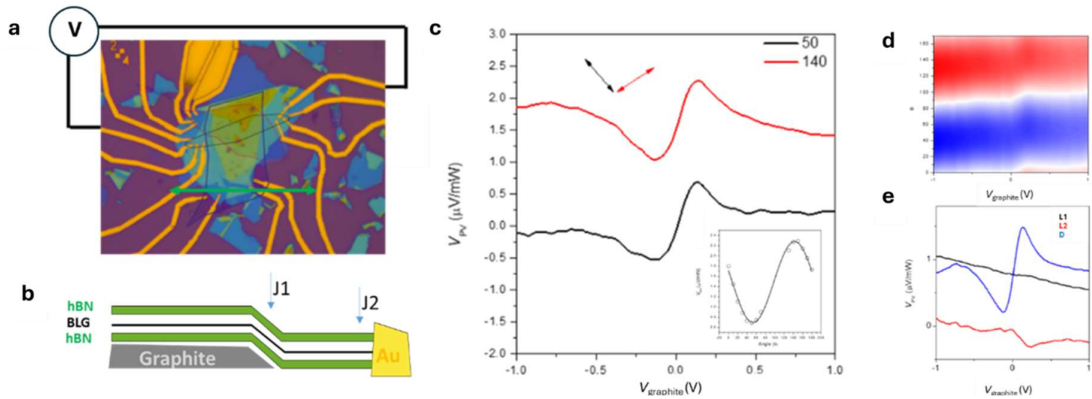

**Figure S6 Control measurements in Bilayer graphene. a,** Optical image of the studied device. Coloured triangles highlight the edges of bilayer graphene and top/bottom graphite gates. **b,** Schematic depicts the heterostructure and the presence of extrinsic junctions induced by gated/un-gated regions of the channel (J1) and local doping by gold contacts (J2). **c,** Photovoltage $V_{PV}$ measured as a function of the bottom gate voltage $V_{Graphite}$ plotted for two orthogonal (black and red) polarization directions. Inset: Polarization dependence plotted for a fixed gate voltage $V_{Graphite}$ = 0.2 V. Open circles are experimental data and solid line is a sinusoidal fit. **d,** polarizatioxn dependent component of the photoresponse plotted as a function of polarization direction ($\theta_p$) and $V_{Graphite}$. **e,** Coefficients obtained from the fits required to describe the entire polarization dependent response. Black, red and blue plot L1, L2 and the offset D respectively.

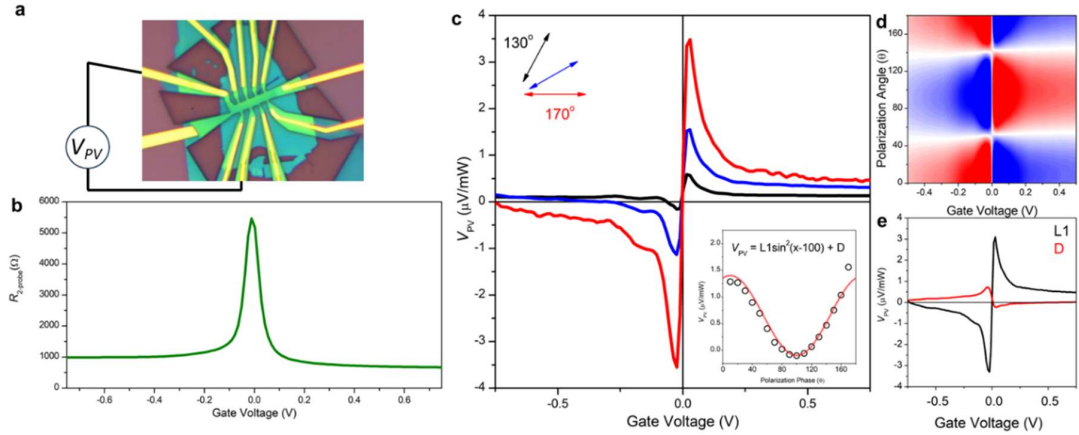

**Figure S7 Control measurements in single layer graphene. a,** Optical image of the studied device with the voltage probes used for measurements fixed. **b,** two-probe resistance ($R_{2\text{-probe}}$) as a function of gate voltage measured between the same pair of contacts depicted in **a**. **c,** Photovoltage ($V_{PV}$) measured as a function of gate voltage for three different polarization directions (depicted by coloured arrows). Inset: Full polarization dependence for a fixed gate voltage $\sim 0.1$ V. Open circles plot experimental data and solid red line plots a sinusoidal fit, specifically a $\sin^2$ function. **d,** polarization dependent component of the photoresponse $V_{Lin}$ plotted as a function of the polarization direction ($\theta_p$) and gate voltage ($V_{Graphite}$). **e,** Extracted coefficients from the fits to the $\sin^2$ function, L1 and the offset D. All measurements are performed at 10 K.

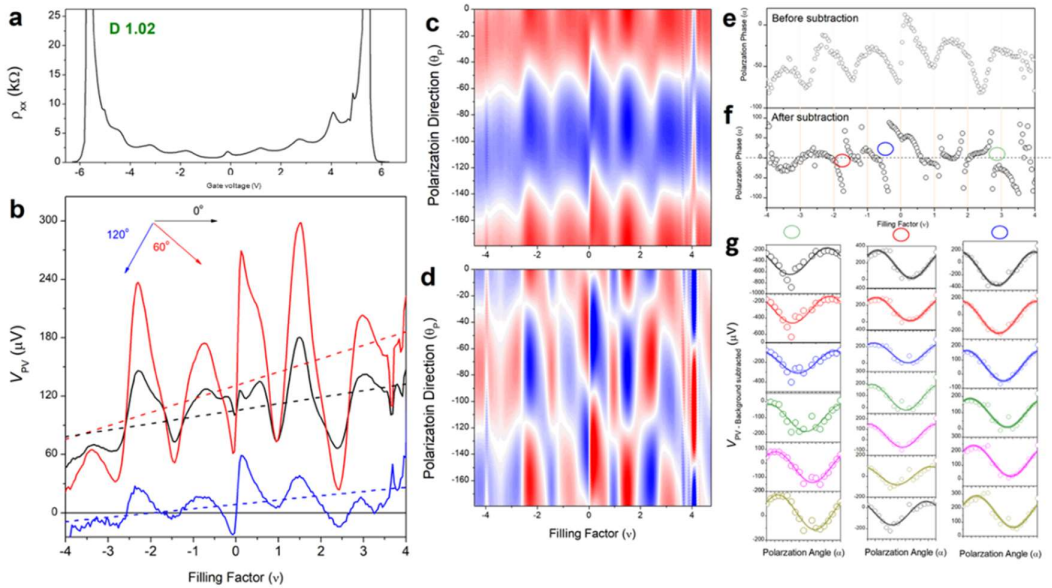

**Figure S8 Measurements in D1.02. a,** Resistivity ($\rho_{xx}$) as a function of gate volage measured in D1.02 at 10 K. **b,** photovoltage ($V_{PV}$) as a function of filling factor ($\nu$) plotted for three polarization directions. The solid lines are experimental data and the dashed lines are linear fits which define the background. **c,** Polarization-dependent component of the photoresponse ($V_{Lin}$) plotted as a function of polarization direction ($\theta_P$) and filling factor ($\nu$). **d,** same as in **c** but after subtracting the linear backgrounds. **e,f** Polarization phases extracted before (**e**) and after (**f**) background subtraction. **g,** $\Delta V_{PV}(\theta_P)$ plotted for different filling factor illustrating the phase drifting behaviours that can be discerned after background subtraction. The three columns illustrate the doping ranges highlighted by coloured circles on **f**.

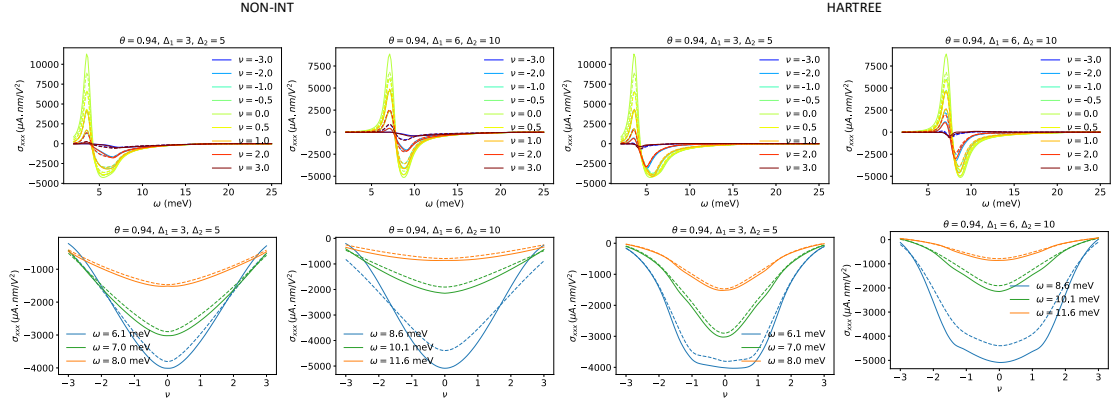

**Figure S9.1:** The $\sigma_{xxx}$ component of shift-current conductivity as a function of frequency (upper panel) and filling (lower panel) for TBG at twist angle $\theta = 0.94°$. The two left columns depict the non-interacting case for different sets of layer-dependent sublattice offset ($\Delta_1$, $\Delta_2$) at T=5 K (solid lines) and T=20 K (dashed lines). On the right side, we added the Hartree contributions from electron-electron interactions. We notice that Hartree corrections do not impact the $\sigma_{xxx}$ component significantly.

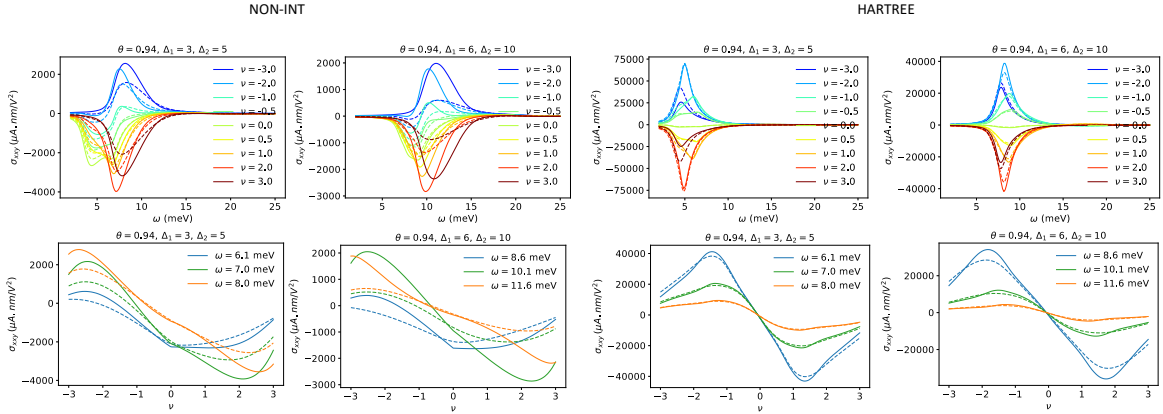

**Figure S9.2:** The $\sigma_{xxy}$ component of shift-current conductivity as a function of frequency (upper panel) and filling (lower panel) for TBG at twist angle $\theta = 0.94°$. The two left columns depict the non-interacting case for different sets of layer-dependent sublattice offset ($\Delta_1$, $\Delta_2$) at T=5 K (solid lines) and T=20 K (dashed lines). On the right side, we added the Hartree contributions from electron-electron interactions. We notice that Hartree corrections impact the $\sigma_{xxy}$ component significantly.

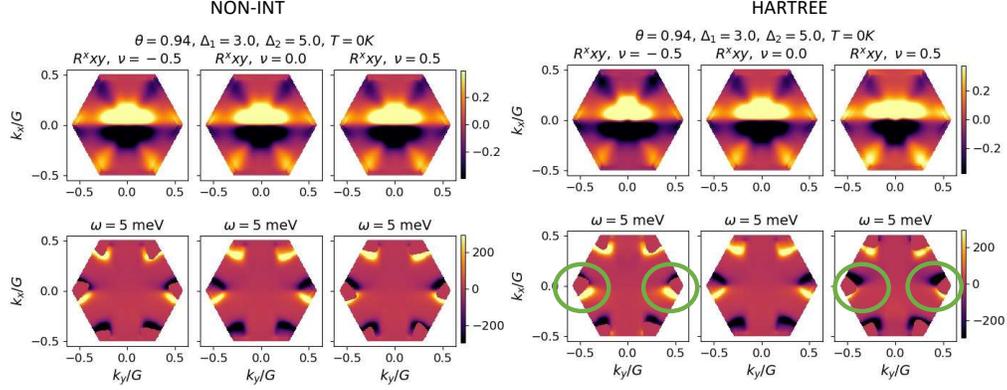

**Figure S9.3:** Shift-vector integrand for $\sigma_{xxy}$ component for the non-interacting case (left panels) and with electron-electron interactions (right panels). The upper rows show the full integrand without the joint density of states at three different fillings of -0.5, 0, and 0.5. The lower rows include the joint density of states at frequency, $\omega = 5\ meV$. We notice that the $\sigma_{xxy}$ component has almost equal regions of positive and negative values of shift-current integrand at zero filling. For the non-interacting case, a small but opposite imbalance in these two regions can be seen at finite filling of $\nu = \pm 0.5$. Interaction-induced effects significantly enhance this imbalance, as shown by expanding yellow and blue regions in green circles.

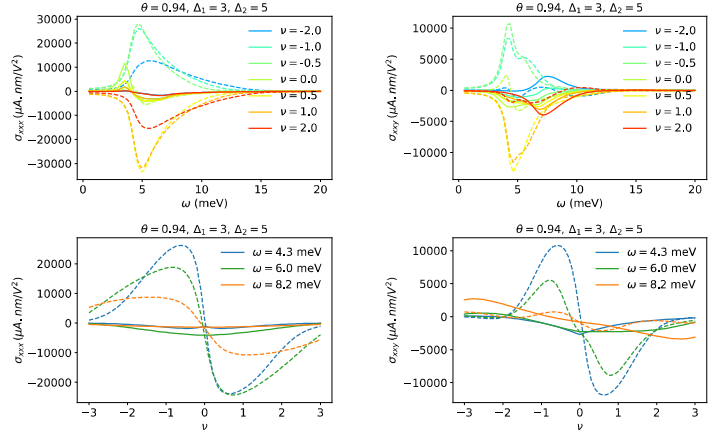

**Figure S9.4:** The $\sigma_{xxx}$ and $\sigma_{xxy}$ component of shift-current conductivity as a function of frequency (upper panel) and filling (lower panel) for TBG at twist angle $\theta = 0.94°$ with strain $\epsilon = 0.1\%$ (dashed line) and without strain (solid line)at T=5 K. We notice that strain significantly enhances both the components. In particular, the $\sigma_{xxx}$ component becomes very large and changes sign with the sign change in filling values.

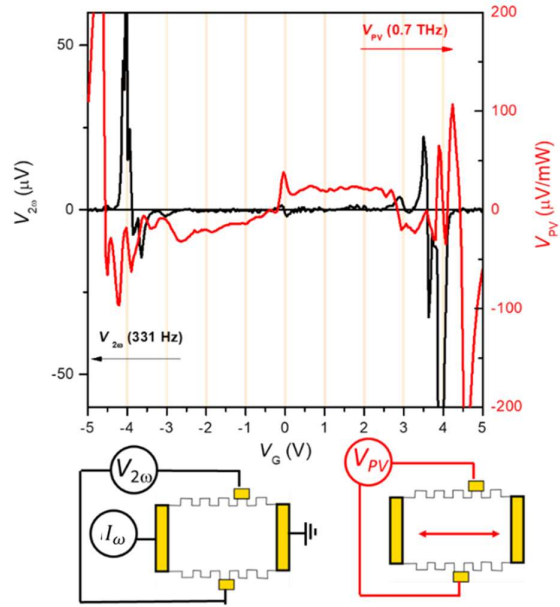

**Figure. S10 Measurement of second-harmonic response via AC quantum transport measurements.** Left axis: Voltage response ($V_{2\omega}$) measured at twice the frequency of AC excitation ($I_{\omega = 331\,Hz}$) for different doping levels ($V_G$ = gate voltage). Right axis (red): Photovoltage $V_{PV}$ measured at 0.7 THz excitation with the polarization direction fixed in the same direction as the current drive measured in quantum transport (left axis). The schematics for both measurements are illustrated in black and red for quantum transport and photovoltage measurements respectively.

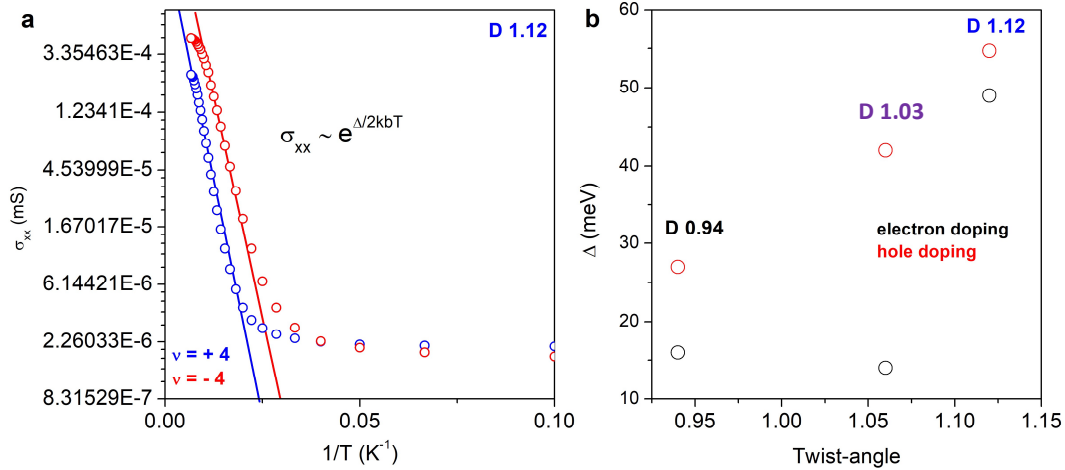

**Figure S11. Band gap extraction via quantum transport. a,** Conductivity ($\sigma_{xx}$) plotted as a function of inverse temperature $T^{-1}$ for fixed doping in the insulating states for electron (blue) and hole (red) doping in one of the studied devices. The open circles are experimental data, and the solid lines are fits to the exponential dependence from which the gaps ($\Delta$) can be extracted. **b,** The band gaps separating flat and remote bands are extracted experimentally in the three studied samples of the main text, with the red and black circles plotting the gaps for hole and electron doping respectively.

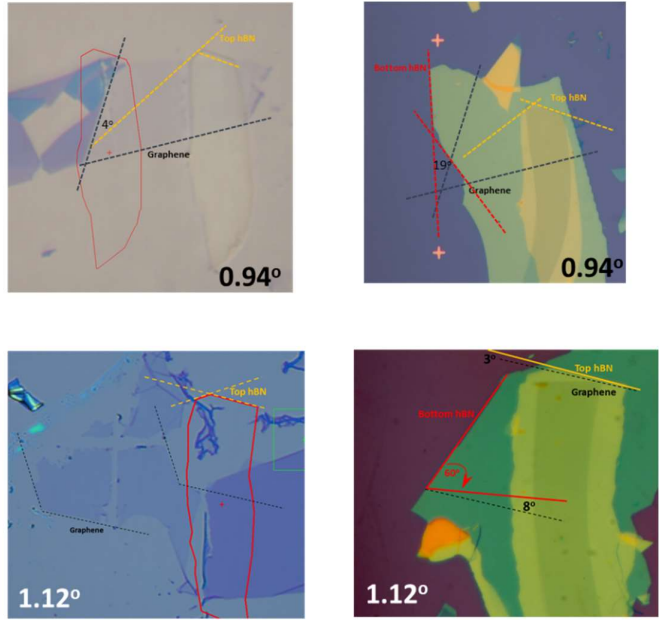

**Figure. S12 Angle alignment between straight edges of graphene and hexagonal boron nitride in D0.94 and D1.12 magic-angle devices.** Optical images of the heterostructures used in this study during the assembly process. The yellow , red and black lines traces top hexagonal-boron nitride (hBN), bottom hBN and graphene flakes respectively.

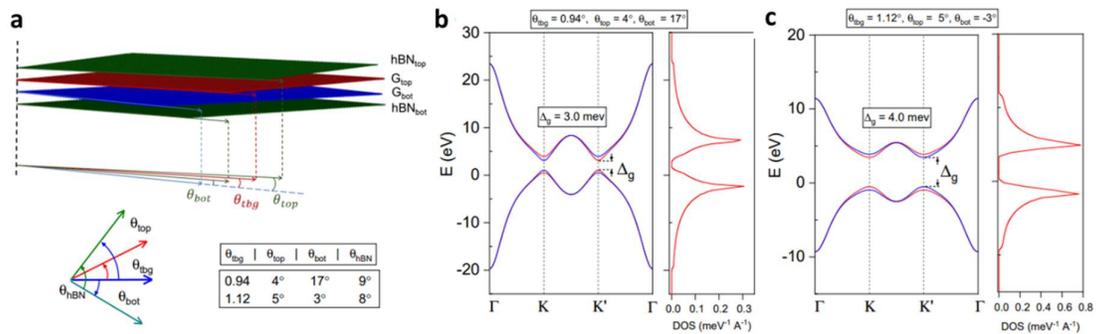

**Figure S13 tight-binding calculations of TBG on hBN. a,** schematic of the modelled TBG heterostructure. The Table indicates the measured alignment angles used for the calculations which are equal to or comparable to those extracted from optical images (Fig. S11). **b, c** calculated band structures for the heterostructures zoomed around the lowest lying bands. Pronounced gaps can be found ($\Delta$) separating them. Right panels in **b** and **c** plot the density of states (DOS).

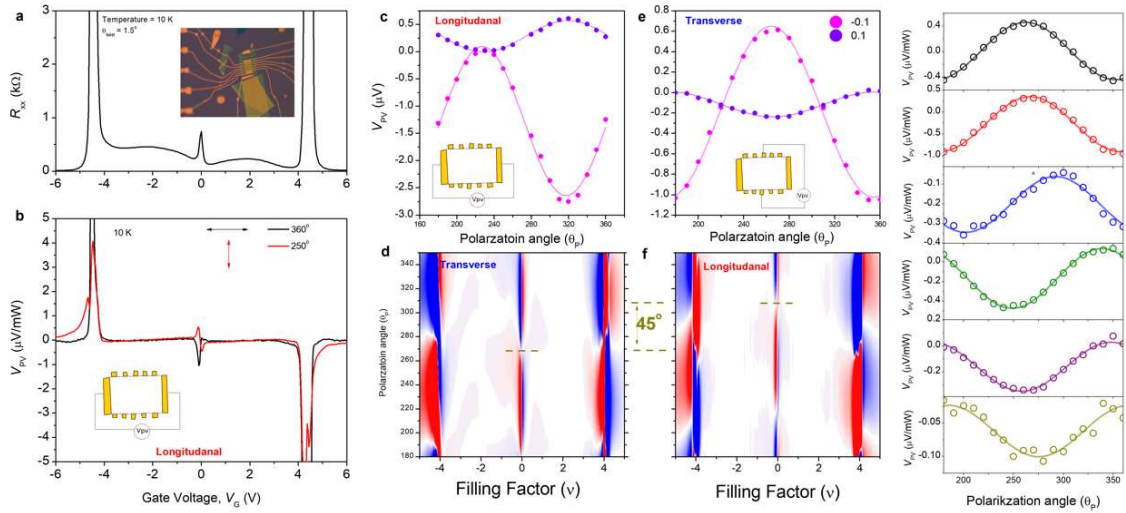

**Figure S14 Photocurrent measurements in D1.5 and proof of C3 symmetry. a,** resistivity ($R_{xx}$) measured as a function of gate voltage ($V_G$) in D1.05 at 10 K. Inset: optical image of the device. **b,** photovoltage ($V_{PV}$) measured at 2.5 THz (10 meV) excitation for two orthogonal polarization directions indicated by black and red arrows. Inset plots a schematic of the device mesa and measurement geometry. **c,e,** $V_{PV}$ measured as a function of polarization angle ($\theta_P$) in two different geometries for the same doping levels. **d,f,** polarization-dependent component of the photoresponse $V_{Lin}$ plotted for the same two geometries as a function of $\theta_P$ and filling factor ($\nu$). The yellow dashed lines indicate the polarization phase in the vicinity of the CNP. **g,** polarization dependence of the photovoltage ($V_{PV}$) measured in the vicinity of the CNP. The open circles are experimental data and the solid lines sinusoidal fits.

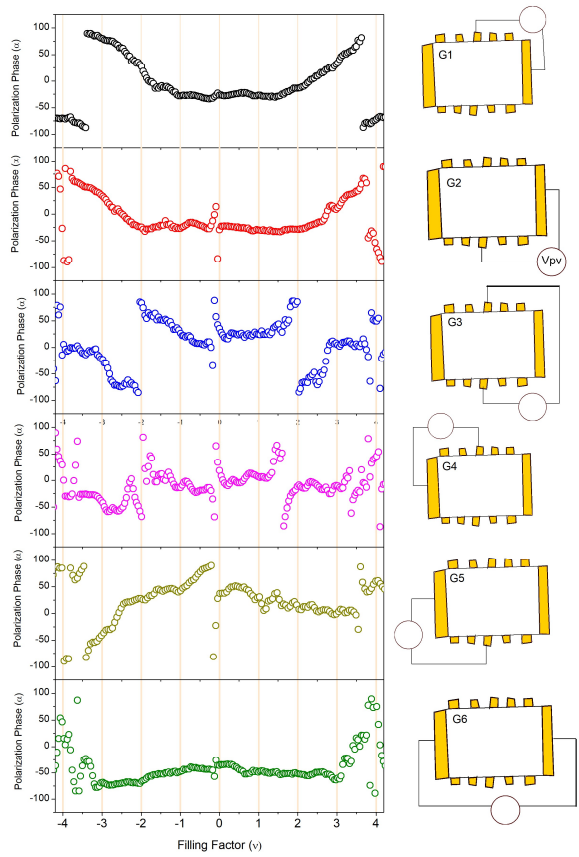

**Supplementary Figure S15 Measurements between different contact pairs in D1.12** Polarization phase (α) measured as a function of filling factor (ν) between different contact pairs in D1.12 at a frequency of 0.7 THz (2.7 meV). Plots on the left are the extracted polarization phase for each of the different measurement geometries (G1-6) sketched on the right-hand side.

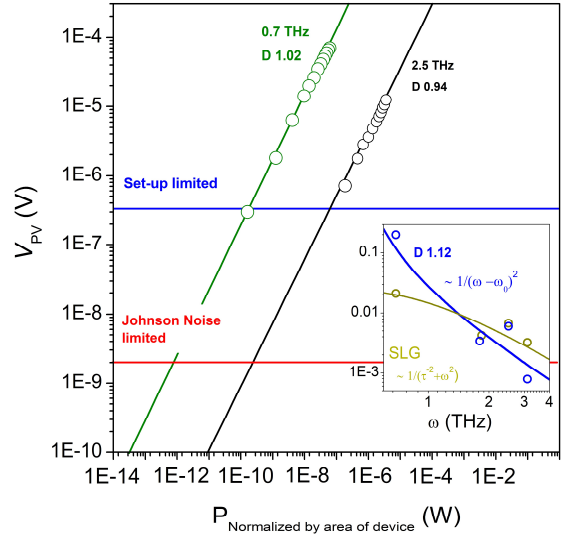

**Figure. S16 THz photodetection with magic-angle twisted bilayer graphene.** Photovoltage ($V_{pv}$) as a function of power normalized by the device size relative to the diffraction-limited spot size ($P_{normalized\ by\ area\ of\ device}$) for fixed filling factor $\nu$ = - 0.5 in D1.02 (green) at 0.7 THz and D0.94 at 2.5 THz. Open circles are experimental data and solid lines are linear fits. The dark blue and dark red lines represent the set-up limited and intrinsic device limited (thermal) noise-floors from which the estimated NEPs are extracted (see methods). **Inset:** wavelength dependence of the photovoltage measured in D1.12 at 0.7 THz for optimal doping $\nu$ = - 3.5 (blue open circles) and fixed $\theta_p$ = 50°. The blue solid line plots function $V_{PV}$ = a/$(\omega - \omega_0)^2$ (solid red line) where $a$ = 1 x $10^{-5}$ and $\omega_0$ = 0.4 THz. The green circles plot data measured in a SLG device for optimal doping and the sold green line plots a fit to a Drude-like response[32], $V_{PV}$ = a/$(\tau^{-2}+\omega^2)$  dependence with a = 2.83 x $10^{-5}$  and $\tau$ = 1.02 x $10^{-12}$ s.

**Supplementary References**

1.  Xu, X., Gabor, N. M., Alden, J. S., Van Der Zande, A. M. & McEuen, P. L. Photo-thermoelectric effect at a graphene interface junction. *Nano Lett.* **10**, 562–566 (2010).

2.  Gabor, N. M. *et al.* Hot carrier-assisted intrinsic photoresponse in graphene. *Science (80-. ).* **334**, 648–652 (2011).

3.  Novelli, P., Torre, I., Koppens, F. H. L., Taddei, F. & Polini, M. Optical and plasmonic properties of twisted bilayer graphene: Impact of interlayer tunneling asymmetry and ground-state charge inhomogeneity. *Phys. Rev. B* **102**, 125403 (2020).

4.  Castilla, S. *et al.* Fast and Sensitive Terahertz Detection Using an Antenna-Integrated Graphene pn Junction. *Nano Lett.* **19**, 2765–2773 (2019).

5.  Bandurin, D. A. *et al.* Resonant terahertz detection using graphene plasmons. *Nat. Commun.* **9**, 5392 (2018).

6.  Cutler, M. & Mott, N. F. Observation of Anderson Localization in an Electron Gas. *Phys. Rev.* **181**, 1336–1340 (1969).

7.  Chaudhary, S., Lewandowski, C. & Refael, G. Shift-current response as a probe of quantum geometry and electron-electron interactions in twisted bilayer graphene.

8.  Kaplan, D., Holder, T. & Yan, B. Twisted photovoltaics at terahertz frequencies from momentum shift current. *Phys. Rev. Res.* **4**, (2022).

9.  Ma, C. *et al.* Intelligent infrared sensing enabled by tunable moiré quantum geometry. *Nature* **604**, 266–272 (2022).

10. Candussio, S. *et al.* Edge photocurrent driven by terahertz electric field in bilayer graphene. *Phys. Rev. B* **102**, (2020).

11. Tielrooij, K. J. *et al.* Hot-carrier photocurrent effects at graphene-metal interfaces. *J. Phys. Condens. Matter* **27**, 164207 (2015).

12. Wei, J. *et al.* Zero-bias mid-infrared graphene photodetectors with bulk photoresponse and calibration-free polarization detection. *Nat. Commun.* **11**, 6404 (2020).

13. Semkin, V. A. *et al.* Zero-Bias Photodetection in 2D Materials via Geometric Design of Contacts. *Nano Lett.* **23**, 5250–5256 (2023).

14. Xu, X., Gabor, N. M., Alden, J. S., van der Zande, A. M. & McEuen, P. L. Photo-Thermoelectric Effect at a Graphene Interface Junction. *Nano Lett.* **10**, 562–566 (2010).

15. Kim, K. *et al.* Tunable moiré bands and strong correlations in small-twist-angle bilayer graphene. *Proc. Natl. Acad. Sci. U. S. A.* **114**, 3364–3369 (2017).

16. Bi, Z., Yuan, N. F. Q. & Fu, L. Designing flat bands by strain. *Phys. Rev. B* **100**, 35448 (2019).

17. Cosma, D. A., Wallbank, J. R., Cheianov, V. & Fal'ko, V. I. Moiré pattern as a

magnifying glass for strain and dislocations in van der Waals heterostructures. *Faraday Discuss.* **173**, 137–143 (2014).

18.  Cea, T., Pantaleón, P. A. & Guinea, F. Band structure of twisted bilayer graphene on hexagonal boron nitride. *Phys. Rev. B* **102**, 155136 (2020).

19.  Sharpe, A. L. *et al. Emergent ferromagnetism near three-quarters filling in twisted bilayer graphene*. https://arxiv.org/pdf/1901.03520.pdf (2019).

20.  Serlin, M. *et al.* Intrinsic quantized anomalous Hall effect in a moiré heterostructure. *Science (80-. ).* **367**, 900–903 (2020).

21.  Niels C.H. Hesp, Sergi Batlle-Porro, Roshan Krishna Kumar, Hitesh Agarwal, David Barcons-Ruiz, Hanan Herzig Sheinfux, Kenji Watanabe, Takashi Taniguchi, Petr Stepanov, F. H. L. K. Cryogenic nano-imaging of second-order moiré superlattices. *arXiv2302.05487v1 [.*

22.  Long, M. *et al.* An atomistic approach for the structural and electronic properties of twisted bilayer graphene-boron nitride heterostructures. *npj Comput. Mater.* **8**, 73 (2022).

23.  Plimpton, S. Fast Parallel Algorithms for Short-Range Molecular Dynamics. *J. Comput. Phys.* **117**, 1–19 (1995).

24.  O'Connor, T. C., Andzelm, J. & Robbins, M. O. AIREBO-M: A reactive model for hydrocarbons at extreme pressures. *J. Chem. Phys.* **142**, 24903 (2015).

25.  Maaravi, T., Leven, I., Azuri, I., Kronik, L. & Hod, O. Interlayer Potential for Homogeneous Graphene and Hexagonal Boron Nitride Systems: Reparametrization for Many-Body Dispersion Effects. *J. Phys. Chem. C* **121**, 22826–22835 (2017).

26.  Slotman, G. J. *et al.* Effect of Structural Relaxation on the Electronic Structure of Graphene on Hexagonal Boron Nitride. *Phys. Rev. Lett.* **115**, 186801 (2015).

27.  Long, M. *et al.* Electronic properties of twisted bilayer graphene suspended and encapsulated with hexagonal boron nitride. *Phys. Rev. B* **107**, 115140 (2023).

28.  Wallbank, J. R., Patel, A. A., Mucha-Kruczyński, M., Geim, A. K. & Fal'ko, V. I. Generic miniband structure of graphene on a hexagonal substrate. *Phys. Rev. B* **87**, 245408 (2013).

29.  Cea, T. & Guinea, F. Band structure and insulating states driven by Coulomb interaction in twisted bilayer graphene. *Phys. Rev. B* **102**, 45107 (2020).

30.  Woods, C. R. *et al.* Commensurate–incommensurate transition in graphene on hexagonal boron nitride. *Nat. Phys.* **10**, 451–456 (2014).

31.  Grover, S. *et al.* Chern mosaic and Berry-curvature magnetism in magic-angle graphene. *Nat. Phys.* **18**, 885–892 (2022).

32.  Mak, K. F. *et al.* Measurement of the optical conductivity of graphene. *Phys. Rev. Lett.* **101**, 196405 (2008).